\documentclass[12pt]{iopart}
\usepackage{bm}
\usepackage{eucal}
\usepackage[normalem]{ulem}
\usepackage[usenames,dvipsnames]{color}
\usepackage{graphicx}
\usepackage{units}
\usepackage{dsfont}
\usepackage[numbers,sort&compress]{natbib}

\usepackage{amsfonts}

\usepackage{etoolbox} 

\newcommand*{\VEC}[1]{\boldsymbol{#1}}
\newcommand*{\TENSOR}[1]{\boldsymbol{\mathsf{#1}}}

\newcommand*{\kB}{k_{\mathrm{B}}}

\newcommand{\D}{\mathcal{D}}
\newcommand{\sign}{\mathop{\mathrm{sign}}}

\newcommand*{\Da}{D_{\mathrm{a}}}
\newcommand*{\ta}{\tau_{\mathrm{a}}}
\newcommand*{\traj}[1]{\underline{#1}}
\newcommand*{\Traj}[1]{\overline{#1}}

\newcommand*{\pp}[1]{\mathfrak{p}[{#1}]}

\newcommand{\T}{\mathsf{T}}

\newcommand{\<}{\left\langle}	
\renewcommand{\>}{\right\rangle}	
\newcommand{\lmat}{\left( \begin{array}{cc}}	
\newcommand{\rmat}{\end{array} \right)}		

\newcommand{\ifl}{\mathrm{ifl}}
\newcommand{\jmp}{\mathrm{jmp}}

\makeatletter
\def\@mkboth#1#2{}
\newlength\appendixwidth
\preto\appendix{\addtocontents{toc}{\protect\patchl@section}}
\newcommand{\patchl@section}{%
  \settowidth{\appendixwidth}{\textbf{Appendix }}%
  \addtolength{\appendixwidth}{1.5em}%
  \patchcmd{\l@section}{1.5em}{\appendixwidth}{}{\ddt}%
}
\makeatother

\begin{document}

\title{How irreversible are steady-state trajectories of a trapped active particle?}
\author{Lennart Dabelow$^1$, Stefano Bo$^2$ and Ralf Eichhorn$^3$}
\address{$^1$Fakult\"at f\"ur Physik, Universit\"at Bielefeld, 33615 Bielefeld, Germany}
\address{$^2$Max-Planck-Institute for the Physics of Complex Systems, N\"othnitzerstr. 38, 01187 Dresden, Germany}
\address{$^3$Nordita, Royal Institute of Technology and Stockholm University,
Roslagstullsbacken 23, SE-106 91 Stockholm, Sweden}
\eads{ldabelow@physik.uni-bielefeld.de, stefabo@pks.mpg.de, ralf.eichhorn@nordita.org}
\begin{abstract}
The defining feature of active particles is that they constantly propel themselves by locally converting chemical energy
into directed motion. This active self-propulsion prevents them from equilibrating
with their thermal environment (e.g., an aqueous solution), thus keeping them permanently out of equilibrium.
Nevertheless, the spatial dynamics of active particles might share certain equilibrium features,
in particular in the steady state.
We here focus on the time-reversal symmetry
of individual spatial trajectories as a distinct equilibrium characteristic.
We investigate to what extent the steady-state trajectories of
a trapped active particle obey or break this time-reversal symmetry.
Within the framework of active Ornstein-Uhlenbeck particles
we find that the steady-state trajectories
in a harmonic potential
fulfill path-wise time-reversal symmetry exactly,
while this symmetry is typically broken in anharmonic potentials.
\end{abstract}
\tableofcontents
\date{\today}
\maketitle

\section{Introduction}
For ordinary, \emph{passive} Brownian motion in a confining potential,
the steady state
is in thermal equilibrium with the surrounding heat bath.
Accordingly, the steady-state dynamics is reversible, i.e.\
it is equally likely to observe a specific trajectory being
traced out forward in time as to observe it being traced out
in the reversed direction backward in time.
This path-wise reversibility is ultimately connected to
conservation of entropy.
Along any trajectory, entropy production in the environment 
is exactly canceled by the entropy change in the system.
These statements can be made mathematically precise by comparing
path probability densities for the forward and backward dynamics
\cite{Jarzynski:2011eai,Seifert:2012stf,Seifert:2005epa}.

For \emph{active} Brownian motion
\cite{romanczuk2012active,elgeti2015physics,Bechinger:2016api,ramaswamy2017active,fodor2018statistical},
on the other hand, the active self-propulsion drive
creates a perpetual non-equilibrium situation which persists even
in the steady state in a confining potential.
Typical systems we have in mind are active colloids or living bacteria in
suspension \cite{elgeti2015physics,Bechinger:2016api,aranson2013active},
which are manipulated by optical tweezers \cite{Argun:2016nbs}.
The active particles are maintained out of equilibrium by the
microscopic processes generating the active self-propulsion.
The details behind these processes are, however, mostly
irrelevant for the dynamical and collective behavior emerging
on the scales of the size of the active particles.
Moreover, they are
often inaccessible in typical experimental setups.
Current video microscopy technology is generally unable to map out the
microscopic details of, e.g., the movement of flagella of a bacterium
or to track the chemical reactions on the surface of a Janus-colloid,
let alone to separate the induced active motion entirely from thermal fluctuations.
Individual trajectories of particles (in the form of particle position as a
function of time) are therefore the central physical observables in a typical experiment.
The question then arises,
in how far the irreversible non-equilibrium nature of the active self-propulsion
is visible (or detectable) on the level of individual particle trajectories
\cite{Fodor:2016hff,nardini2017entropy,dabelow2019irreversibility,caprini2019entropy,
flenner2020active,dabelow2020}.
Answering this question contributes to the endeavor of developing
a theoretical framework for the ``thermodynamics of active matter''
\cite{ginot2015nonequilibrium,solon2015pressure,takatori2015towards,preisler2016configurational,wittmann2017effective,solon2018generalized,solon2018generalizedNJP,paliwal2018chemical,pietzonka2019autonomous,mandal2019motility,crosato2019irreversibility,dal2019linear,grandpre2020entropy,wexler2020dynamics,fodor2020dissipation}
as it helps to understand
under which conditions the (collective) steady state of active matter emerging
from self-propulsion appears to have equilibrium characteristics and
under which conditions
its non-equilibrium nature becomes apparent.

We here focus on the trajectory-wise (ir)reversibility as described in the
beginning of the Introduction
to assess how closely active steady-state dynamics resemble equilibrium.
In order to exclude the possibility of
activity-induced currents in the spatial coordinate
as obvious signatures of non-equilibrium,
we consider the simple situation of active Brownian motion in
a time-independent one-dimensional confining potential with active driving which
is unbiased in space and time.
Since the microscopic mechanisms which generate the active self-propulsion
are of no interest for our trajectory-based analysis,
we adopt the common strategy to include self-propulsion as an effective force
into the equations of motion for a Brownian particle \cite{romanczuk2012active,Bechinger:2016api}.
A particularly successful such model is the so-called
active Ornstein-Uhlenbeck particle (AOUP) \cite{martin2020statistical}.
In the AOUP model the active force is represented by a
stochastic Ornstein-Uhlenbeck process
\cite{Gardiner:HandbookOfStochasticMethods,vanKampen1992stochastic},
without including the matching damping kernel \cite{zamponi2005fluctuation}
such that the equilibrium fluctuation-dissipation relation \cite{Kubo:1966tfd,marconi2008fluctuation}
is not fulfilled.
In that way, the AOUP is constantly
driven out of thermal equilibrium by a fluctuating ``self-propulsion'' force
with exponentially decaying correlations 
which represent the directional persistence typical for active self-propulsion.
Due to its conceptual simplicity, the AOUP has become 
a quite popular and successful model for active Brownian motion
\cite{Fily:2012aps,Farage:2015eii, 
Maggi:2014gee,Argun:2016nbs,maggi2017memory, 
Fodor:2016hff,marconi2017heat,mandal2017entropy,Puglisi:2017crf,
koumakis2014directed,szamel2014self,szamel2015glassy,maggi2015multidimensional,Flenner:2016tng,paoluzzi2016critical,
Marconi:2016vdi,szamel2017evaluating,sandford2017pressure,wittmann2017effective,caprini2018linear,fodor2018statistical,bonilla2019active,
woillez2020nonlocal, 
berthier2013non,Marconi:2015tas,shankar2018hidden,dal2019linear,fodor2020dissipation}. 

We present the mathematical details of the AOUP model in the
next Section. Then, in Section~\ref{sec:irr} we introduce the statistical
weight of individual trajectories as our main quantity of interest
in order to quantify irreversibility by comparing forward and
backward paths. We study different ``minimal'' trapping potentials in
Section~\ref{sec:HP} (harmonic trap), Section~\ref{sec:DWP} (anharmonic double-well),
and Section~\ref{sec:QP} (anharmonic single-well)
with surprising findings concerning the (ir)reversibility
of individual steady-state trajectories of the AOUP in the various traps.
We conclude in Section~\ref{sec:conclusions} with a short summary
and discussion.

\section{Model}
\label{sec:model}
We study a single particle in one dimension, which is in contact with a thermal environment
at temperature $T$ and, additionally, is driven by active Ornstein-Uhlenbeck fluctuations.
This so-called active Ornstein-Uhlenbeck particle (AOUP)
is confined by an external potential $U(x)$ with
$U(x) \to \infty$ as $|x| \to \infty$ (faster than logarithmically).
The overdamped equation of motion reads
\begin{equation}
\label{eq:xLE}
	\dot{x}(t) = -\frac{1}{\gamma} U'(x(t)) + \sqrt{2\Da} \, \eta(t) + \sqrt{2D} \, \xi(t)
\, .
\end{equation}
Here, $\xi(t)$ is a Gaussian white-noise process
with $\<\xi(t)\> = 0$ and $\<\xi(t) \xi(t')\> = \delta(t-t')$.
The thermal diffusion coefficient $D=\kB T/\gamma$
is related to temperature $T$ and viscous friction $\gamma$
via the fluctuation-dissipation relation ($\kB$ is Boltzmann's constant),
indicating that the environment constitutes a thermal bath at equilibrium.
The active fluctuations are described by the stationary Ornstein-Uhlenbeck process $\eta(t)$ satisfying
\begin{equation}
\label{eq:etaLE}
	\dot{\eta}(t) = -\frac{1}{\ta} \eta(t) + \frac{1}{\ta} \zeta(t) 
\end{equation}
with another Gaussian white noise source $\zeta(t)$, which is independent of $\xi(t)$.
Consequently, $\eta(t)$ is a Gaussian process with $\<\eta(t)\> = 0$ and
\begin{equation}
\label{eq:<etaeta>}
	\<\eta(t) \eta(t')\> = \frac{1}{2\ta} \e^{-| t - t' | / \ta}
\, .
\end{equation}
This model of the active fluctuations is not directly related to
the operational details of the self-propulsion drive.
The variable $\eta(t)$ rather provides an \emph{effective},
bath-like description of the drive's statistical properties.
The actual observable quantities in a typical experiment
are particle positions $x(t)$ as a function of time, i.e.\
AOUP trajectories $\Traj{x}=\{x(t)\}_{t=0}^\tau$ starting at time $t=0$
and ending at time $t=\tau$.

\section{Path weights and irreversibility}
\label{sec:irr}
Our central quantity of interest is the probability to observe 
a trajectory $\Traj{x}=\{x(t)\}_{t=0}^\tau$ starting in $x_0$ at time $t=0$.
We calculate this so-called path weight $\pp{\Traj{x}}$ by considering the joint path weight
$\pp{\traj{x},\traj{\eta}|x_0,\eta_0}$
for the combined trajectory $(\traj{x},\traj{\eta})=\{(x(t),\eta(t))\}_{t>0}^\tau$
conditioned on a fixed starting configuration $(x_0,\eta_0)$,
and integrating over all possible realizations of $\Traj{\eta}=\eta_0 \cup \traj{\eta}$,
\begin{equation}
\label{eq:PathWeight}
\pp{\Traj{x}}
	= \int \D\Traj{\eta} \; \pp{\traj{x},\traj{\eta}|x_0,\eta_0} \,
	  p_0(x_0,\eta_0)
\, ,
\end{equation}
where the joint distribution of initial configurations $(x_0,\eta_0)$ is given by
$p_0(x_0,\eta_0)$.
Since the combined process $(\traj{x}, \traj{\eta})$ is Markovian, we
can express its path weight via the standard Onsager-Machlup path integral
\cite{Onsager:1953fai,Machlup:1953fai,cugliandolo2017rules}
\begin{equation}
\fl
\label{eq:PathWeightJoint}
\pp{\traj{x},\traj{\eta}|x_0,\eta_0}
	\propto \exp \left\{ -\int_0^\tau \!\! \d t
		\left[ \frac{(\dot{x}_t - v_t - \sqrt{2\Da} \, \eta_t )^2}{4D}
			 + \frac{(\ta \dot{\eta}_t + \eta_t)^2}{2}
			 + \frac{1}{2}\frac{\partial v_t}{\partial x}
		\right] \right\}
\, ,
\end{equation}
where we introduce the shorthand notation
$x_t \equiv x(t)$, $\eta_t \equiv \eta(t)$ and
$v_t \equiv -U'(x(t))/\gamma$.
Note that the integral in \eref{eq:PathWeightJoint} has been derived using a mid-point
discretization (Stratonovich convention), which we tacitly assume for all stochastic integrals
in the following.
This path weight
is quadratic in the active driving $\eta(t)$, such that
the functional integral $\int\D\Traj{\eta}$ in \eref{eq:PathWeight} is
a ``Gaussian integral in function space'' and can be performed analytically
\cite{dabelow2019irreversibility},
provided that the initial distribution is
Gaussian in $\eta_0$ too, because $\int\D\Traj{\eta}$ includes the
integral over the initial point $\eta_0$.

Using the path weight $\pp{\Traj{x}}$, 
we can then assess the irreversibility of individual trajectories by comparing the probability
$\pp{\Traj{x}}$ to observe a trajectory $\Traj{x} = \{x(t)\}_{t=0}^\tau$
forward in time with the probability $\pp{\Traj{\tilde{x}}}$ to observe
its time-reversed twin $\Traj{\tilde{x}} = \{x(\tau-t)\}_{t=0}^\tau$
\footnote{In the absence of time-dependent forces, which is the case in our current setting,
the probability functional $\mathfrak p$ is the same for forward and backward processes.
Only if there were a time-dependent experimental protocol for the external forces,
we would need to reverse this protocol as well and would in general obtain a different functional
$\tilde{\mathfrak p}$.}.
In fact, the log ratio of path probabilities
\begin{equation}
\label{eq:DeltaSigmaDef}
\Delta\Sigma[\Traj{x}] = \kB \ln \frac{\pp{\Traj{x}}}{\pp{\Traj{\tilde{x}}}}
\end{equation}
has been adopted as a natural measure to quantify the ``breaking''
of time-reversal symmetry, i.e.\ to quantify irreversibility
\cite{Jarzynski:2011eai,Seifert:2012stf,seifert2018stochastic}.
By definition, the process appears reversible if $\Delta\Sigma[\Traj{x}] = 0$,
since then the forward and backward trajectories occur with identical probabilities.
A positive value indicates that the trajectory $\Traj{x}$ is more likely to be observed
than its time-reversed twin $\Traj{\tilde{x}}$ and vice versa for a negative value.
Its mean (averaged over all possible trajectories) provides a way to estimate the arrow of time
on statistical grounds \cite{roldan2015decision}.
Physically, the log ratio of path probabilities~\eref{eq:DeltaSigmaDef}
can be related to the entropy production in the thermal environment and the
path-wise mutual information between particle trajectories and active fluctuations \cite{dabelow2019irreversibility}.
Without activity ($\Da = 0$), i.e.\ for passive particles, 
it coincides with the standard notion of total entropy production
as defined in stochastic thermodynamics~\cite{Seifert:2005epa,Seifert:2012stf}.

\begin{figure}
\centering
\includegraphics[scale=1]{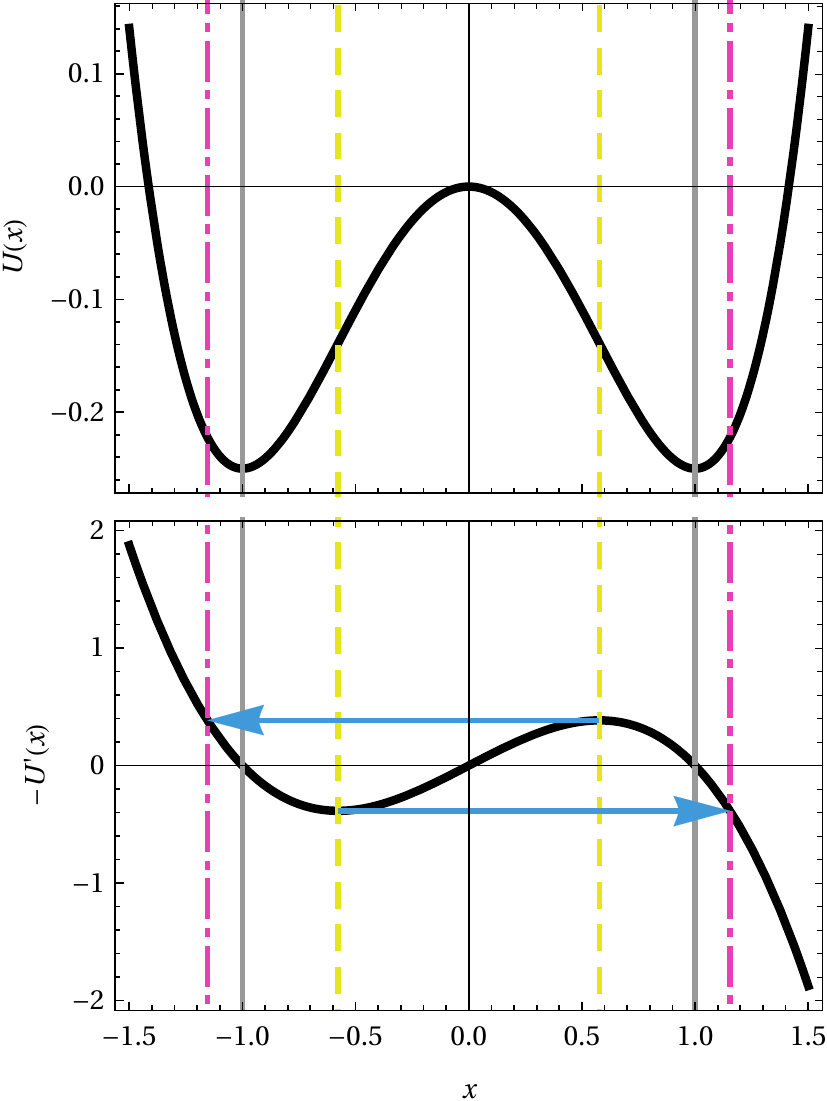}
\caption{Anharmonic potential $U(x)$ from~\eref{eq:V} for $k_4 = 1$ and $k_2 = -1$ (upper panel)
and corresponding force $-U'(x)$ (lower panel).
Potential minima are marked by solid gray lines.
Inflection points are marked by dashed yellow lines.
The corresponding conjugate points of equal force are marked by dash-dotted purple lines.
The blue arrows illustrate the jumps in the high-persistence limit of large $\ta$.
}
\label{fig:Potential}
\end{figure}
In the following we apply $\Delta\Sigma[\Traj{x}]$ for assessing
the reversible or irreversible character of
steady-state trajectories of an AOUP
trapped in a confining potential.
It will turn out that the simplest trapping potentials
to reveal essential irreversibility features
are a quartic potential ($k_4 \geq 0$),
\begin{equation}
\label{eq:V}
	U(x) = \frac{k_4}{4} x^4 + \frac{k_2}{2} x^2
\, ,
\end{equation}
and its various ``special cases'', i.e.\
a harmonic trap ($k_4=0$ and $k_2>0$, Section~\ref{sec:HP}),
a double-well potential ($k_4>0$ and $k_2<0$, Section~\ref{sec:DWP})
and a quartic single-well potential ($k_4>0$ and $k_2=0$, Section~\ref{sec:QP}).
See Fig.~\ref{fig:Potential} (upper panel) for an example of a quartic double-well
potential with $k_2=-1$ and $k_4=1$ in \eref{eq:V}.
We are interested in trajectories that start, remain and end
in the steady state.
However, in a typical initial setup (short) transients are often unavoidable,
for instance, because the initial particle distribution did not yet
relax within the potential or did not yet build
up its steady-state correlations with the active fluctuations.
The relative importance of these transients quickly diminishes over time
and thus they are irrelevant for
sufficiently long trajectories, i.e.\ longer than a few relaxation times
within the potential and correlation times of the active fluctuations.
In other words, performing numerical experiments
for long enough trajectories, the effect of short initial transients
while approaching the joint steady state of $x$ and $\eta$ can be neglected.

\section{Harmonic trap}
\label{sec:HP}
In this Section, we consider the case $k_4=0$ and $k_2>0$ in \eref{eq:V}.
This case of a harmonic trap is interesting, because,
by extending our
method from \cite{dabelow2019irreversibility},
we can calculate $\Delta\Sigma[\Traj{x}]$ analytically for an
initial distribution $p_0(x_0,\eta_0)$ in \eref{eq:PathWeight}
representing the joint steady state.
We thus end up with a steady-state expression for $\Delta\Sigma[\Traj{x}]$
that does not contain any of the above mentioned transients.

\subsection{Path probability ratio in the joint steady state}
The joint system with degrees of freedom $\VEC{q}=(x, \eta)$
is a two-dimensional Ornstein-Uhlenbeck process,
whose steady-state distribution takes the form
\cite{Risken:TheFokkerPlanckEquation,Gardiner:HandbookOfStochasticMethods}
\begin{equation}
\label{eq:pSS}
p_\infty(x,\eta)
	= \frac{\e^{ -\frac{1}{2} \, \VEC{q}^\T \TENSOR{C}^{-1} \VEC{q} } }
		   {\sqrt{(2\pi)^2 \, \det \TENSOR{C}}}
\, ,
\end{equation}
where
\begin{equation}
\label{eq:C}
\TENSOR{C}
= \lmat
	\frac{\gamma}{k_2} \left[ D + \Da \rho \right] & \sqrt{\frac{\Da}{2}} \rho  \\
	\sqrt{\frac{\Da}{2}} \rho  & \frac{1}{2 \ta}
\rmat
\, , \qquad 
\rho = \frac{1}{1 + \frac{k_2 \, \ta}{\gamma}}
\, .
\end{equation}
Taking \eref{eq:pSS} as the initial distribution,
$p_0(x_0,\eta_0)=p_\infty(x_0,\eta_0)$, we can write the path weight
\eref{eq:PathWeight} as (see also \eref{eq:PathWeightJoint})
\begin{eqnarray}
\label{eq:PathWeightHP}
\pp{\Traj{x}} & \propto & \int \D\Traj{\eta} \; p_\infty(x_0,\eta_0) \,
\exp \int_{0}^\tau \d t \; \Bigg\{
	- \frac{1}{4D} \left[ \dot{x}_t - v_t - \sqrt{2\Da} \, \eta_t \right]^2
\nonumber \\
&&\quad\qquad\qquad \mbox{}
	- \frac{\ta^2}{2} \left[ \dot{\eta}_t + \frac{1}{\ta} \eta_t \right]^2 
	- \frac{1}{2} \frac{\partial v_t}{\partial x}
\Bigg\}
\nonumber \\
& = & \int \D\Traj{\eta} \, \exp \Bigg\{ \int_0^\tau \d t \,
	\left[ -\frac{\left( \dot{x}_t - v_t \right)^2}{4D} 
		   - \frac{1}{2} \frac{\partial v_t}{\partial x}
	\right]
	- \frac{1}{2} \bar{C}_{11} x_0^2
\nonumber \\
&& \quad\qquad\qquad \mbox{}
+ \int_0^\tau \d t \,
	\left[ \frac{\sqrt{2\Da}}{2D} \left( \dot{x}_t - v_t \right) - \delta(t) \bar{C}_{12} x_t \right] \eta_t
\nonumber \\
&& \quad\qquad\qquad \mbox{}
	- \frac{1}{2} \int_0^\tau \d t \int_0^\tau \d t^\prime \; \eta_t V_{\mathrm{HP}}(t, t^\prime) \eta_{t^\prime}
\Bigg\}
\end{eqnarray} 
after partial integration of the $\dot{\eta}_t$ terms in the second equality,
and with the abbreviation
\begin{equation}
\label{eq:Cbar}
\fl
\bar{\TENSOR{C}} =
\left( \begin{array}{cc}
\bar{C}_{11} & \bar{C}_{12} \\
\bar{C}_{12} & \bar{C}_{22}
\end{array} \right)
= \TENSOR{C}^{-1}
=
\frac{1}{D+\Da\rho^2}
\left( \begin{array}{cc}
k_2/\gamma 									& -\sqrt{2\Da}\,\frac{k_2\ta}{\gamma}\rho \\
-\sqrt{2\Da}\,\frac{k_2\ta}{\gamma}\rho 	& 2\ta(D+\Da\rho)
\end{array} \right)
\end{equation}
for the (symmetric) inverse of $\TENSOR{C}$.
The differential operator $V_{\mathrm{HP}}(t, t^\prime)$ reads
\begin{eqnarray}
\label{eq:VHP}
\fl
V_{\mathrm{HP}}(t, t')
	= \delta(t - t') \left[
		- \ta^2 \partial_{t'}^2 + 1 + \frac{\Da}{D}
		+ \delta(t') \left( -\ta^2 \partial_{t'} - \ta + \bar{C}_{22} \right)
		+ \delta(\tau - t') \left(\ta^2 \partial_{t'} + \ta \right)
	\right]
\, .
\nonumber \\
\end{eqnarray}
Performing the Gaussian integral over $\Traj{\eta}$  we obtain
\begin{eqnarray}
\label{eq:PathWeightPositionHP}
\fl
\pp{\Traj{x}}
\propto \exp \left\{ \int_0^\tau \d t \int_0^\tau \d t^\prime
	\left[ -\frac{1}{4D}
		\left( \dot{x}_t + \frac{k_2}{\gamma} x_t \right)
		\left[ \delta(t-t^\prime) - \frac{D_a}{D} \Gamma_{\mathrm{HP}}(t, t^\prime) \right]
		\left( \dot{x}_{t^\prime} + \frac{k_2}{\gamma} x_{t^\prime} \right)
	\right]
\right.
\nonumber \\
\qquad\qquad\qquad \left. \mbox{}
	- \int_0^\tau \d t \;
		\frac{\sqrt{2\Da}}{2D} \bar{C}_{12} \Gamma_{\mathrm{HP}}(t, 0)
		\left( \dot{x}_t + \frac{k_2}{\gamma} x_t \right) x_0
\right.
\nonumber \\
\qquad\qquad\qquad \left. \mbox{}
	- \frac{1}{2} \left[ \bar{C}_{11} - \bar{C}_{12}^{2} \, \Gamma_{\mathrm{HP}}(0,0) \right] x_0^2
\right\}
\, ,
\end{eqnarray}
where we have plugged in $v_t=-k_2 x_t/\gamma$. Moreover,
$\Gamma_{\mathrm{HP}}(t, t')$ denotes the operator inverse
of $V_{\mathrm{HP}}(t, t')$ in the sense that
$\int_0^\tau \d t' \; V_{\mathrm{HP}}(t, t') \Gamma_{\mathrm{HP}}(t', t'') = \delta(t - t'')$
(since the operator $V_{\mathrm{HP}}(t, t')$ is diagonal in $t$ and $t'$,
the integral simplifies into a differential equation with $\Gamma_{\mathrm{HP}}(t', t'')$
being its Green's function, see \ref{app:GammaHP}).
It can be constructed by a suitable extension of the procedure in \cite{dabelow2019irreversibility}
(see~\ref{app:GammaHP}),
\begin{eqnarray}
\label{eq:GammaHP}
\fl
\Gamma_{\mathrm{HP}}(t, t')
= \frac{\kappa_{+-} \e^{-\lambda |t - t'|} + \kappa_{-+} \e^{-\lambda(2\tau - |t - t'|)}
		- \kappa_{++} \e^{-\lambda(t+t')} - \kappa_{--} \e^{-\lambda(2\tau - t - t')}}
	   {2 \ta^2 \lambda \left( \kappa_{+-} - \kappa_{-+} \e^{-2 \lambda\tau} \right)}
\, .
\end{eqnarray}
with the abbreviations
\begin{equation}
\label{eq:kappapmpm:def}
\kappa_{\pm\pm} = 
\kappa_\pm \left( 1 - \kappa_\pm\ta/\bar{C}_{22} \right)
\, , \quad
\kappa_{\pm} = 1 \pm \sqrt{1 + \Da / D}
\, ,
\end{equation}
and $\lambda = \sqrt{1 + \Da/D} / \ta$.
The first subscript in $\kappa_{\pm\pm}$ refers to the first $\kappa_\pm$
on the right-hand side of its definition and the second
subscript to the second $\kappa_\pm$.

Expression \eref{eq:PathWeightPositionHP} together with the Green's function
$\Gamma_{\mathrm{HP}}(t, t^\prime)$ is thus the path probability density for a Brownian
particle trapped in a harmonic potential and driven by an active Ornstein-Uhlenbeck process,
when the two start out in a joint steady state.
To compare to the case of independent initial conditions, which we have studied
in \cite{dabelow2019irreversibility}
and which we briefly re-capitulate in Section~\ref{sec:DWP} below,
we observe that
in this case $\bar{C}_{12} = 0$ and $\bar{C}_{22} = 2\ta$,
so that we recover \eref{eq:GammaGen} as stated below
when using $1-\kappa_\pm/2 = \kappa_\mp/2$.

The path probability density for the \emph{time-reversed} trajectories $\Traj{\tilde{x}}$
is given by the same expression \eref{eq:PathWeightPositionHP} (just equipping
all $x$-symbols with a tilde), because there is no external driving protocol,
and the system remains in its stationary state at all times along
the forward trajectory so that the initial condition
for the backward trajectory
is again the steady-state distribution \eref{eq:pSS}.
Using $\tilde{x}(t)=x(\tau-t)$ and $\tilde{x}_0=x(\tau)=x_\tau$,
we can then express $\pp{\Traj{\tilde{x}}}$ in terms of the forward path. 
The resulting expression looks similar to \eref{eq:PathWeightPositionHP},
but with opposite sign for all $\dot{x}_t$-terms, $x_0$ substituted by $x_\tau$,
and the replacements $t \to \tau-t$, $t' \to \tau-t'$ in all
non-trivial time-arguments of $\Gamma_{\mathrm{HP}}$.
\newpage
With these results we can finally calculate $\Delta\Sigma[\Traj{x}]$
as defined in \eref{eq:DeltaSigmaDef},
\begin{eqnarray}
\label{eq:DeltaSigmaHP}
\fl
\frac{\Delta\Sigma[\Traj{x}]}{\kB} =
\frac{1}{D} \int_0^\tau \d t \int_0^\tau \d t^\prime
\left\{ 
	\dot{x}_t \, \left(-\frac{k_2}{\gamma} x_{t^\prime}\right)
	\left[ \delta(t-t^\prime) - \frac{\Da}{D} \bar{\Gamma}_{\mathrm{HP}}(t, t^\prime) \right]
\right.
\nonumber \\
\qquad\qquad\qquad \left. \mbox{}
  + \frac{\Da}{4D}\dot{x}_t \dot{x}_{t^\prime} \Delta\Gamma_{\mathrm{HP}}(t, t^\prime)
  + \frac{\Da}{4D}\left(\frac{k_2}{\gamma}\right)^{\!\!2} x_t x_{t^\prime} \Delta\Gamma_{\mathrm{HP}}(t, t^\prime)
\right\}
\nonumber \\
\qquad \mbox{}
	- \frac{\sqrt{2\Da}}{2D} \int_0^\tau \d t \, \Bigg\{
		\dot{x}_t \, \bar{C}_{12}
			\Big[ \Gamma_{\mathrm{HP}}(t, 0)x_0 + \Gamma_{\mathrm{HP}}(\tau-t, 0)x_\tau \Big]
\nonumber \\
\qquad\qquad\qquad\qquad \left. \mbox{}
	   + \frac{k_2}{\gamma} x_t \, \bar{C}_{12}
			\Big[ \Gamma_{\mathrm{HP}}(t, 0)x_0 - \Gamma_{\mathrm{HP}}(\tau-t, 0)x_\tau \Big]
	\right\}
\nonumber \\
\qquad  \mbox{}
	+ \frac{1}{2} \left[ \bar{C}_{11} - \bar{C}_{12}^{2} \, \Gamma_{\mathrm{HP}}(0,0) \right]
	       	      \left( x_\tau^2 - x_0^2 \right)
\, .
\end{eqnarray}
To arrive at this expression we have used the symmetry
$\Gamma_{\mathrm{HP}}(t, t^\prime)=\Gamma_{\mathrm{HP}}(t^\prime, t)$,
and we have introduced the abbreviations
$\bar{\Gamma}_{\mathrm{HP}}=\frac{1}{2}\left[ \Gamma_{\mathrm{HP}}(t, t^\prime)+\Gamma_{\mathrm{HP}}(\tau-t, \tau-t^\prime)\right]$
and
$\Delta\Gamma_{\mathrm{HP}}=\Gamma_{\mathrm{HP}}(t, t^\prime)-\Gamma_{\mathrm{HP}}(\tau-t, \tau-t^\prime)$
for the mean and the difference of $\Gamma_{\mathrm{HP}}(t, t^\prime)$ and its time-reversed counterpart
$\Gamma_{\mathrm{HP}}(\tau-t, \tau-t^\prime)$.

\subsection{Path-wise reversibility}
Substituting the inverse \eref{eq:Cbar} of~\eref{eq:C} into~\eref{eq:DeltaSigmaHP},
we can show that $\Delta\Sigma[\Traj{x}]=0$, and thus, according to \eref{eq:DeltaSigmaDef},
\begin{equation}
\label{eq:HPreversible}
	\frac{ \pp{\Traj{\tilde{x}}} }{ \pp{\Traj{x}} } = 1
\end{equation}
exactly, for any values of the system parameters.
This reversibility holds on the level of individual steady-state trajectories of
arbitrary duration $\tau$.
Transients (towards the steady state) may still be irreversible, but are not captured
in \eref{eq:HPreversible} by construction, since we are interested in characterizing
the (ir)reversibility of the steady state of a trapped AOUP and therefore
calculated $\Gamma_{\mathrm{HP}}(t,t')$ for an initial setup
corresponding to the joint steady state of the Brownian particle and the active
Ornstein-Uhlenbeck fluctuations.

The details of the derivation of \eref{eq:HPreversible} involve some rather tedious manipulations
of the log ratio~\eref{eq:DeltaSigmaHP} and are thus
relegated to~\ref{app:DeltaSigmaHP=0}.
The crucial insight behind these calculations
is to integrate by parts in order to move all time derivatives
from the particle trajectory to the memory kernel $\Gamma_{\mathrm{HP}}(t,t^\prime)$,
generating additional terms with one or two time points on the boundary $t=0$
or $t=\tau$.
Then we can demonstrate that all three types
of contributions to~\eref{eq:DeltaSigmaHP},
namely the two-time integrals over $t$ and $t'$,
the one-time integrals with the other time point lying on the boundary,
and the boundary contributions (which include the integral involving $\delta(t-t')$),
vanish individually.

The ratio~\eref{eq:HPreversible} tells us that for the steady state
in a harmonic trap we will observe every individual trajectory
with a probability which is exactly equal to the probability for observing the
same trajectory in reversed time, i.e.\
probabilities to observe forward and backward paths coincide and the process
appears reversible.
In a corresponding experiment in which only the position of the particle
is measured, it is therefore impossible to determine an ``arrow of time''.
Given a movie of the particle dynamics, we will be unable to tell whether
it is being played forwards or backwards, no matter how long it is, i.e.\
no matter how many data points we have.
As argued above, the unbiased, time-correlated fluctuations do not
favor a particular direction in space or time,
so we may have expected this result. 
In the following Sections we will check if this
reversibility property is specific to the combination of
an Ornstein-Uhlenbeck noise modeling
the active forces with a harmonic trapping potential,
or if it is valid for more general confining potentials.

Before doing so, we remark that the situation is different in a scenario
in which the active fluctuations
are an observable degree of freedom.
If we were somehow able to measure the force $\sqrt{2 \Da}\,\eta(t)$ stemming
from the active fluctuations, we could compare probabilities for forward and
backward histories of the \emph{joint process} $(x(t), \eta(t))$.
In this $x$-$\eta$ phase space, there is always an effective ``torque''
(a nonconservative force component) as can be seen from the fact that $\eta(t)$ drives
$\dot x(t)$, but there is no feedback from $x(t)$ to $\dot\eta(t)$
(see Eqs.~\eref{eq:xLE} and \eref{eq:etaLE}).
Hence, there is a ``current'' in the $(x(t),\eta(t))$ dynamics in the form of a net average rotation,
breaking the phase space symmetry, which will allow to distinguish forward from backward
processes on statistical grounds
(see, e.g., Ref.~\cite{dadhichi2018origins} and Fig.~1 therein).
For the particle coordinates alone, a similar situation 
can arise in more than one spatial dimension
under the influence of a mechanical torque
\cite{volpe2006torque,volpe2007brownian,
Filliger:2007bga,chiang2017electrical,argun2017experimental,cerasoli2018asymmetry,pietzonka2018universal}.

\subsection{Remark on $d>1$ dimensions}
It is straightforward to convince ourselves that the same path-wise
reversibility holds for the steady state of an AOUP trapped in a
harmonic potential of higher than one dimension
(or many non-interacting AOUPs in such a potential).
In that case, the equations of motion read
\numparts
\begin{eqnarray}
\label{eq:xLEd}
\dot{\VEC{x}}(t) = -\TENSOR{K}\VEC{x}(t) + \sqrt{2\Da}\,\VEC{\eta}(t) + \sqrt{2D}\,\VEC{\xi}(t)
\, ,
\\
\label{eq:etaLEd}
\dot{\VEC{\eta}}(t) = -\frac{1}{\ta}\VEC{\eta}(t) + \frac{1}{\ta} \VEC{\zeta}(t)
\, ,
\end{eqnarray}
\endnumparts
with vector quantities of dimension $d>1$ as obvious generalizations
of the scalar quantities from \eref{eq:xLE} and \eref{eq:etaLE},
and a positive definite symmetric $d \times d$ tensor $\TENSOR{K}$
specifying the harmonic trap.
Note that the different dimensions (or particles) have identical
properties concerning thermal and active fluctuations, i.e.\
the $d \times d$ tensors in front of $\VEC{\xi}(t)$
and $\VEC{\eta}(t)$ in \eref{eq:xLEd} and $\VEC{\zeta}(t)$ in \eref{eq:etaLEd}
are all proportional to the identity tensor.

Since $\TENSOR{K}$ is symmetric we can diagonalize it with an
orthogonal $d \times d$ tensor $\TENSOR{O}$.
The rotated thermal and active noise processes
$\TENSOR{O}\VEC{\xi}(t)$ and $\TENSOR{O}\VEC{\eta}(t)$
have the same statistical properties as the original
ones. In particular, their components are mutually independent, such that
the set of $2d$ equations \eref{eq:xLEd} and \eref{eq:etaLEd}
decouples into $d$ independent pairs of equations of the form
\eref{eq:xLE}, \eref{eq:etaLE}.

As a consequence, we can write the path probability density $\pp{\Traj{\VEC{x}}}$ for
the $d$-dimensional particle trajectories $\Traj{\VEC{x}}$ as a product of $d$ independent
path probability densities for the individual components.
Accordingly, the irreversibility measure $\Delta\Sigma[\Traj{\VEC{x}}]$ is
a sum of $d$ independent contributions of the form \eref{eq:DeltaSigmaHP},
which all vanish identically.

\section{Double-well potential}
\label{sec:DWP}
In this Section, we consider the AOUP dynamics \eref{eq:xLE}
in a double-well potential like the one shown in Fig.~\ref{fig:Potential},
corresponding to choosing $k_4>0$ and $k_2<0$ in \eref{eq:V}.
Unlike the purely harmonic trap from the previous Section, this case does not allow
for calculating $\Delta\Sigma[\Traj{x}]$ with
the joint steady state as initial distribution, because the analytical form
of this joint steady state is unknown.
We therefore use the results from \cite{dabelow2019irreversibility},
which we have obtained
under the assumption that the active fluctuations alone are in their steady state,
independent of the initial position $x_0$, i.e.\
\begin{equation}
\label{eq:p0SteadyEta}
p_0(x_0,\eta_0) = p_0(x_0) p_0(\eta_0|x_0) = p_0(x_0) p_0(\eta_0)
= p_0(x_0) \, \sqrt{\frac{\ta}{\pi}} \e^{-\ta \eta_0^2}
\, ,
\end{equation}
where $\sqrt{\frac{\ta}{\pi}} \e^{-\ta \eta^2}$ is the steady state
distribution of the process \eref{eq:etaLE}.
For reasonable $x_0$, we are thus a short
transient away from the joint steady state, with negligible effects on the long-term steady
state dynamics of the system.

The log ratio of path probabilities from \cite{dabelow2019irreversibility}
for a fixed initial position $x_0$ of the active particle reads
\begin{eqnarray}
\label{eq:DeltaSigmaGen}
\frac{\Delta\Sigma[\Traj{x}]}{\kB}
& = & \ln \frac{\pp{\Traj{x}}}{\pp{\Traj{\tilde{x}}}}
= \ln \frac{\pp{\traj{x}|x_0}}{\pp{\traj{\tilde{x}}|\tilde{x}_0}} + \ln \frac{p(x_0)}{p(\tilde{x}_0)}
\nonumber \\
& = & \frac{1}{D} \int_0^\tau \!\! \d t \int_0^\tau \!\! \d t' \;
	\dot{x}_t \, v_{t'} \left[ \delta(t-t') - \frac{\Da}{D} \Gamma(t,t') \right]
  + \ln \frac{p(x_0)}{p(x_\tau)}
\, ,
\end{eqnarray}
where $v_t = -U'(x(t))/\gamma$ and
\begin{equation}
\label{eq:GammaGen}
\fl
\Gamma(t,t') =
	\left( \frac{1}{2 \ta^2 \lambda} \right)
	\frac{ \kappa_+^2 \e^{-\lambda |t - t'|}
		 + \kappa_-^2 \e^{-\lambda(2\tau - |t - t'|}
		 - \kappa_+ \kappa_- \left[ \e^{-\lambda(t+t')} + \e^{-\lambda(2\tau - t - t')} \right] }
	{ \kappa_+^2 - \kappa_-^2 \e^{-2 \lambda\tau} }
\, ,
\end{equation}
with $\lambda = \sqrt{1 + \Da/D} / \ta$ and $\kappa_{\pm} = 1 \pm \sqrt{1 + \Da/D}$, as before.
The first term in \eref{eq:DeltaSigmaGen} has two contributions in the double integral,
one involving the non-local kernel $\Gamma(t,t')$, which vanishes for $\Da=0$ and is
therefore named the ``colored-noise contribution'',
\numparts
\begin{equation}
\label{eq:DeltaSigmac}
\frac{\Delta\Sigma_{\mathrm{c}}[\Traj{x}]}{\kB}
= -\frac{\Da}{D^2} \int_0^\tau \!\! \d t \int_0^\tau \!\! \d t' \; \dot{x}_t \, v_{t'} \, \Gamma(t,t')
\, ;
\end{equation}
and a term proportional to $\delta(t-t')$,
which survives even for $\Da=0$ and is thus called ``white-noise contribution'',
\begin{equation}
\label{eq:DeltaSigmaw}
\frac{\Delta\Sigma_{\mathrm{w}}[\Traj{x}]}{\kB}
= \frac{1}{D} \int_0^\tau \!\! \d t \; \dot{x}_t \, v_t
= \frac{1}{\kB T} \left[ U(x_0) - U(x_\tau) \right]
\, .
\end{equation}
\endnumparts
The second term in \eref{eq:DeltaSigmaGen} quantifies the irreversibility associated
with the boundary distributions at initial and final points in time,
and is usually interpreted as the change in system entropy
between the initial and final configurations \cite{dabelow2019irreversibility}.
It is non-extensive in the length $\tau$ of the trajectory and
thus turns out to be negligible for sufficiently long trajectories
compared to the time-extensive contributions in $\Delta\Sigma[\Traj{x}]$.

\subsection{First numerical experiments}
\label{sec:firstNumerics}
\begin{figure}
\includegraphics[scale=0.85]{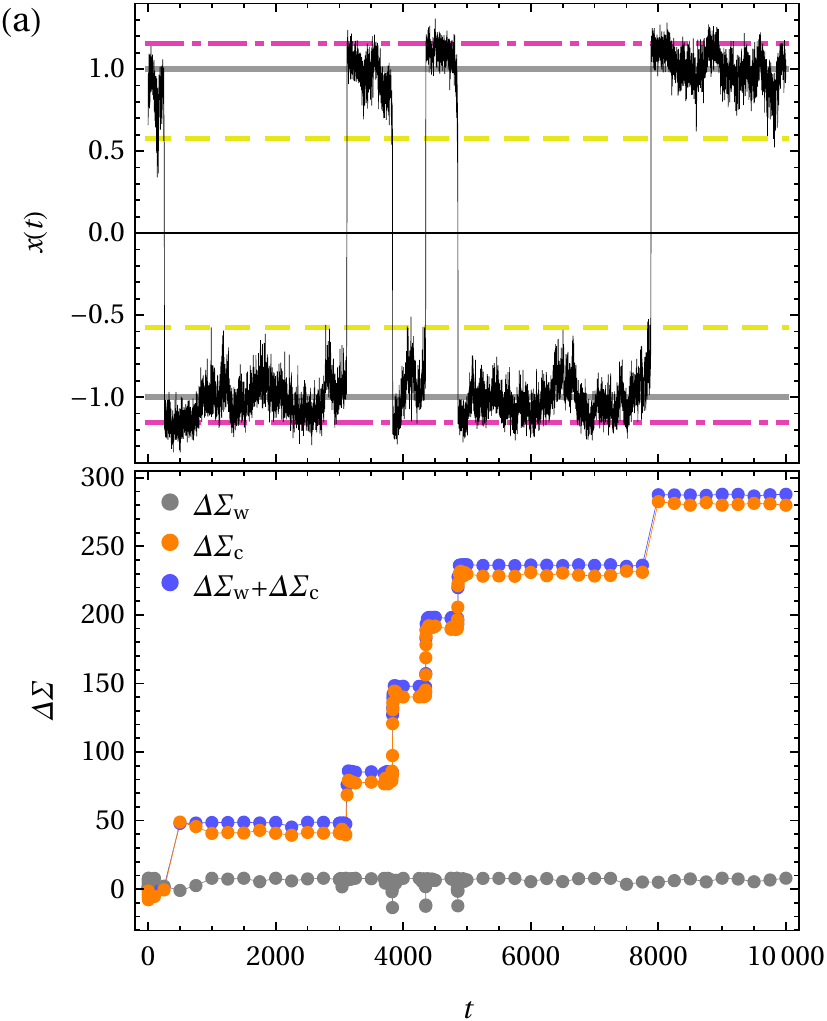}
\hspace*{1cm}
\includegraphics[scale=0.85]{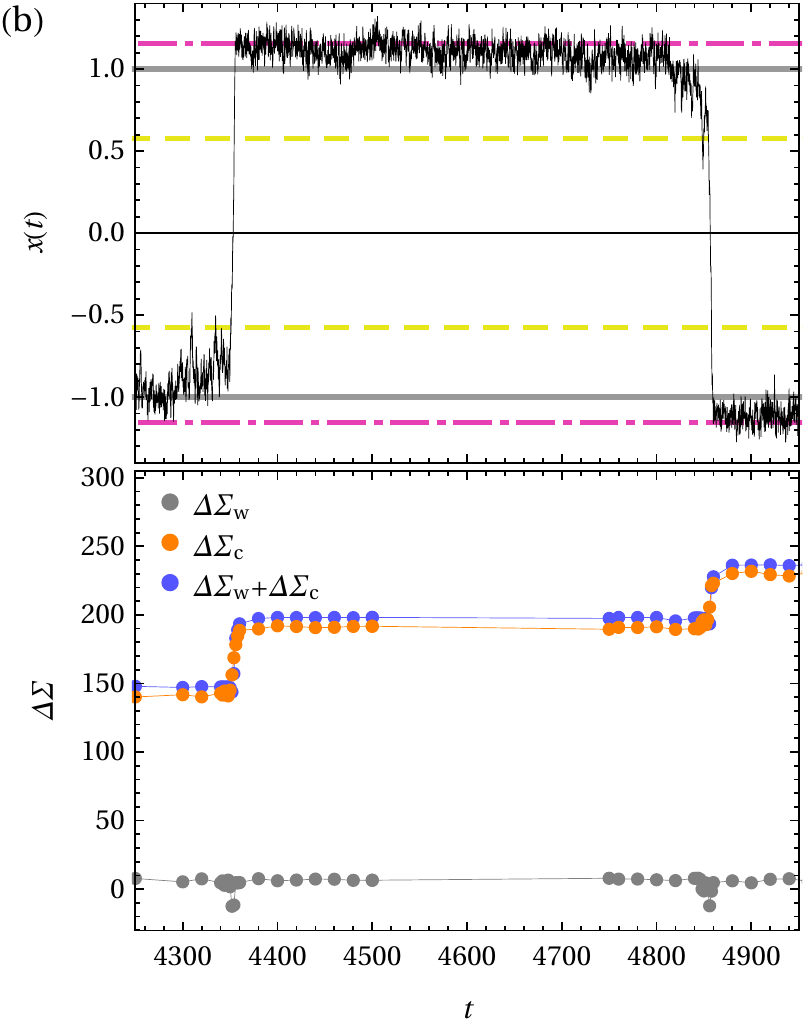}
\caption{(a) A single trajectory (top) and corresponding irreversibility productions $\Delta\Sigma$ (bottom).
(b) Close-up of the same trajectory (top) and irreversibility productions (bottom) for the time interval from about 4300 to 4900.
In the lower panels, the dots mark the time points for which the values of $\Delta\Sigma_{\mathrm{w}}$ (grey dots)
and $\Delta\Sigma_{\mathrm{c}}$ (orange dots) have been evaluated from the integrals in \eref{eq:DeltaSigmaw}
and \eref{eq:DeltaSigmac}; the connecting lines serve as a guide to the eye.
For numerical reasons and since the integrals in \eref{eq:DeltaSigmac} have a
smoothing effect anyway, we chose a time-resolution considerably coarser than the original trajectory.
Yellow dashed and purple dash-dotted lines in the top panels represent the specific positions in the potential
as marked in Fig.~\ref{fig:Potential}, solid gray lines illustrate the positions of the potential minima.
Unlike for a passive particle, for an AOUP the maxima of the steady-state distribution
do not coincide with the potential minima due the active driving.
Parameter values: $k_4 = 1$, $k_2 = -1$, $\ta = 250$, $\Da = 10$, $D = 0.01$, $\gamma = 1$, $\kB=1$.
}
\label{fig:DynIrrev}
\end{figure}
A typical trajectory is plotted in the upper panel of Fig.~\ref{fig:DynIrrev}(a).
The irreversibility $\Delta\Sigma$ as a function of the trajectory duration is shown
in the lower panels (blue),
along with a splitting into white-noise contributions
$\Delta\Sigma_{\mathrm{w}}$ (gray) and colored-noise
contributions $\Delta\Sigma_{\mathrm{c}}$ (orange);
the boundary term $\ln \frac{p(x_0)}{p(\tilde{x}_0)}$
would typically be of the order of $\Delta\Sigma_{\mathrm{w}}$
and has been neglected in the plots.
As it should, the white-noise contribution becomes stationary pretty soon and does not grow
extensively since the corresponding stochastic integral depends only on the initial and final
points of the trajectory.
The active contribution, however, increases sharply whenever the particle jumps from one
minimum to the other.
Hence, we find that the steady-state trajectories in a quartic potential
are irreversible, and that the jumps between the two potential minima render the dynamics irreversible,
even though there is no net current from one side to the other.
This means that it is in principle possible to tell apart trajectories
which run forward in time from those which run backward in time.

\subsection{Explaining irreversibility}
\label{sec:explaining}
The quartic potential for $k_4>0$, $k_2<0$ and the resulting force are sketched in Fig.~\ref{fig:Potential}.
The potential has a local maximum at $x_{\max} = 0$ and two minima at $x_{\min}^\pm = \pm\sqrt{-k_2 / k_4}$
(solid gray lines in Fig.~\ref{fig:Potential}).
There are also two inflection points at $x_{\ifl}^\pm = \pm \sqrt{-k_2 / 3 k_4} = x_{\min} / \sqrt{3}$,
where the force's magnitude reaches a local extremum
(dashed yellow lines in Fig.~\ref{fig:Potential}).
For every inflection point, there is one point on the opposite
side of the origin where the force takes the same value as it does at $x_{\ifl}^\pm$
(dash-dotted purple lines in Fig.~\ref{fig:Potential}).
We denote these conjugate equal-force points by
$x_{\jmp}^\pm = \mp\sqrt{-4 k_2 / 3 k_4} = 2 x_{\ifl}^\mp$,
such that $U'(x_{\jmp}^\pm) = U'(x_{\ifl}^\pm)$.

Fily \cite{fily2019self} analyzed such a setup in the limit of high persistence,
for which $\ta$ is by far the largest time scale in the system
(see also \cite{caprini2019active,woillez2020nonlocal} for related studies).
He showed that, asymptotically for $\ta \to \infty$, the particle is never found
in the region $(x_\ifl^-, x_\ifl^+)$ between the inflection points.
To explain this observation, assume that the potential barrier around the origin is
sufficiently high so that purely thermally induced crossings of the barrier are rare.
However, if the active forcing is large enough to counter the restoring potential forces
when moving uphill towards the barrier,
the particle can climb up the hill towards the
origin and reach the inflection point of maximum counterforce.
Due to persistence of the active fluctuations, the particle will then
continue to push in the direction of its active drive, and speed up because the opposing force will
be weaker until the origin is reached, from where on active and potential
forces will even point in the same direction.
The active forces will only be balanced again once the particle approaches the
conjugate equal-force point beyond the barrier, where it will thus be slowed down and ``stopped'' eventually.
The transition from one side of the double-well potential to the other thus
looks like a ``sudden jump'', in particular on the time-scale $\ta$ of the active fluctuations.
When the active fluctuations change sign, the same process can then occur in the opposite direction.

As a consequence, the trajectories of such an active particle show hysteresis-like behavior
and exhibit a clear signature of an ``arrow of time'' despite the conservative,
time-independent confining forces:
The particle preferantially ``jumps'' from an inflection point $x_{\ifl}^\pm$
(dashed yellow lines in Fig.~\ref{fig:Potential})
to the corresponding equal-force point $x_\jmp^\pm$ on the other side of the barrier
(dash-dotted purple lines in Fig.~\ref{fig:Potential}), but (practically) never the other way
(from $x_{\jmp}^\pm$ to $x_{\ifl}^\pm$).
We can nicely see this behavior in Fig.~\ref{fig:DynIrrev}(b),
showing close-up views of two transitions from the trajectory plotted in Fig.~\ref{fig:DynIrrev}(a).
If we were to see a trajectory dominated by ``jumps'' from the points
$x_{\jmp}^\pm$ to the points $x_{\ifl}^\pm$, we are most likely watching a trajectory
which is ``played'' backward in time.
This intuitive picture suggests that the irreversibility of trajectories is encoded predominantly in the
``jumps'' between the potential wells.
Figure \ref{fig:DynIrrev} highlights that our dynamical measure
$\Delta\Sigma[\Traj{x}]$, eq.~\eref{eq:DeltaSigmaGen},
identifies these ``jumps'' as the principle source of irreversibility, too. 

The irreversibility of trajectories also becomes apparent by inspection of the
two-point probability density $p(x',t';x, t)$, shown for the time delay
$t'-t=25$ in Fig.~\ref{fig:Propagator}.
It illustrates the probability of observing the particle at
a certain position $x$ in the potential at time $t$ and at position $x'$
at a later time $t'=t+25$.
The mirror symmetry $x \leftrightarrow -x$ of the potential is reflected
by the point symmetry in Fig.~\ref{fig:Propagator} with respect to the origin $(0,0)$.
The potential minima are located at the solid gray lines. The bright spots
around the crossings of these lines on the diagonal $x=x'$
(lower left and upper right corner in Fig.~\ref{fig:Propagator})
correspond to a high probability density and indicate that
the particle is most likely 
to stay in the potential well it is starting from
at the beginning of the time-interval $t'-t=25$.
The other two bright spots in Fig.~\ref{fig:Propagator}
(upper left corner and lower right corner) represent the ``jumps''
from one side of the potential to the other side.
Their irreversibility is visible in the asymmetry of the
probability density spots with respect to the
diagonal $x=x'$. The starting positions ($x$ coordinates) at time $t$ are close to the
inflection points $x \approx x_{\ifl}^\pm$ (vertical dashed yellow lines),
while the final positions ($x'$ coordinates) are the jump-target points
$x' \approx x_{\jmp}^\pm$ (horizontal dash-dotted purple lines).
Analogous density spots that would correspond to the opposite ``jumps''
from $x \approx x_{\jmp}^\pm$ to $x' \approx x_{\ifl}^\pm$ are absent,
implying that any steady-state trajectory of the AOUP trapped in a double-well potential
becomes more and more irreversible over time.
In this sense it ``produces irreversibility'' and encodes an
``arrow of time''.
\begin{figure}
\centering
\includegraphics[scale=1.5]{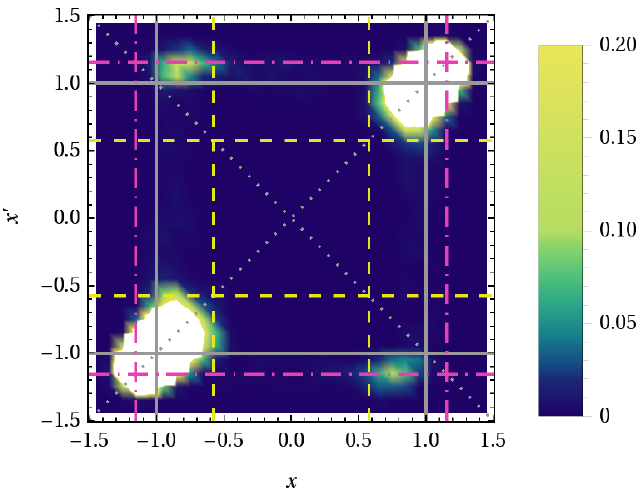}
\caption{Two-point probability density $p(x',t';x,t)$ for step size $t'-t=25$.
Potential minima are marked by solid gray lines.
Inflection points $x_{\ifl}^\pm$ are marked by dashed yellow lines.
The corresponding jump target points $x_\jmp^\pm$ are marked by dash-dotted purple lines,
see also Fig.~\ref{fig:Potential}.
The dotted gray lines mark the diagonals $x=x'$ and $x=-x'$
and guide the eye to assess the (a)symmetry of the plot.
Parameter values as in Fig.~\ref{fig:DynIrrev}.
}
\label{fig:Propagator}
\end{figure}

\subsection{Exploring irreversibility}
As pointed out above, the clear hysteresis-like behavior and
its irreversibility are associated with
the high persistence of the active fluctuations induced by large $\ta$.
We expect, however, that particle trajectories become fully reversible
in the strict limit $\ta \to \infty$, in which
the dynamics of $\eta(t)$ is ``frozen'' and the active velocity
$\sqrt{2\Da}\eta(t)=\sqrt{2\Da}\eta(0)$ remains constant for all times.
In that case the constant active velocity $\sqrt{2\Da}\eta(0)$ can be
absorbed into the potential as a tilting force, $U(x)-\gamma\sqrt{2\Da}\eta(0)\, x$,
such that \eref{eq:xLE} effectively turns into a Langevin-equation for
a \emph{passive} Brownian particle confined in a tilted double-well potential.
The steady state is then identical to thermal equilibrium in the tilted
potential, and therefore
any particle trajectory is reversible
(on the time-scales that are shorter than $\ta$).

For smaller $\ta$-values, on the other hand, we may expect that
a ``weakened'' or ``smoothed-out'' form of the hysteresis-like behavior
is still producing irreversibility, even if the
irreversibility of the trajectories might not be visible to the naked eye any more
in the lucidity it is displayed in Fig.~\ref{fig:DynIrrev}.
To study the effect of decreasing $\ta$, there are two possible scenarios:
On the one hand, we can keep the active diffusion $\Da$ fixed, such that the process
$\sqrt{2\Da} \, \eta(t)$ approaches a thermal white-noise process with
diffusion $\Da$ as $\ta \to 0$.
Then, the particle feels an effective total temperature of
$\gamma (D + \Da)/\kB = T + \gamma\Da/\kB$.
Alternatively, we can require that the active velocity
$v_{\mathrm a} = \sqrt{\Da/\ta}$ remains fixed
(which may be the more natural limit when taking the AOUP as an
approximation for an active Brownian particle \cite{Fily:2012aps,Farage:2015eii,dal2019linear}).
Then, the overall intensity decreases as $\ta$ becomes smaller
until the particle just feels the thermal fluctuations at temperature $\gamma D/\kB=T$ as $\ta \to 0$.
In both cases, the limit $\ta \to 0$ turns \eref{eq:xLE} into an equation of motion
for a \emph{passive} Brownian particle, so that again all trajectories become reversible.
In the following, we will explore the regime of intermediate $\ta$-values for
both of the scenarios mentioned above, namely constant active diffusion
and constant active velocity.
For a discussion of the $\ta \to 0$ and $\ta \to \infty$ limits
for an AOUP in a slightly different setting we refer to Ref.~\cite{dadhichi2018origins}.

\subsubsection{Decreasing active correlation time at constant active diffusion.}
\begin{figure}
\centering
\includegraphics[scale=1.6]{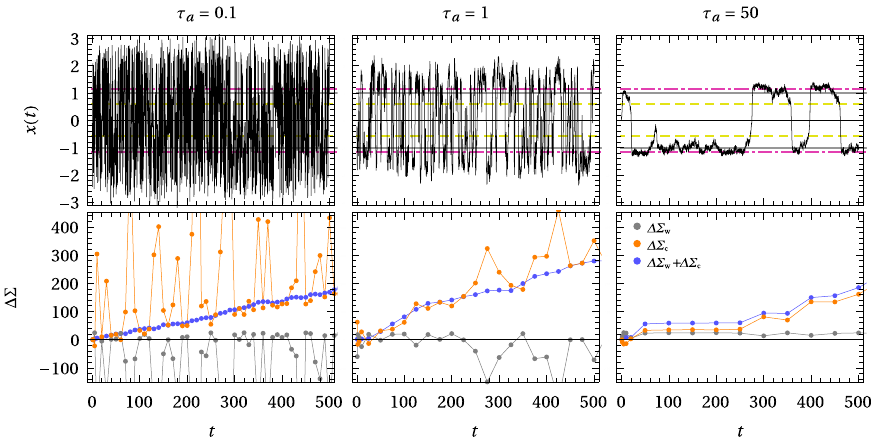}
\caption{Single trajectories (top) and corresponding irreversibility productions $\Delta\Sigma$
(bottom) for three systems with different active correlation times
smaller than $\ta=250$ from Fig.~\ref{fig:DynIrrev}:
$\ta=0.1$ (left), $\ta=1$ (middle), and
$\ta = 50$ (right).
All other parameter values are as in Fig.~\ref{fig:DynIrrev}.
The frequency of jumps increases as $\ta$ decreases and
the self-propulsion force switches direction more often.
In the lower panels, the dots mark the time points for which the values of $\Delta\Sigma_{\mathrm{w}}$
(grey dots) and $\Delta\Sigma_{\mathrm{c}}$ (orange dots)
have been evaluated from the integrals in \eref{eq:DeltaSigmaw}
and \eref{eq:DeltaSigmac}; the connecting lines serve as a guide to the eye.
The time-resolution is considerably lower than for the trajectories in the upper panels.
In particular for smaller values of $\ta$ the increasing number of
jumps in the trajectory is not resolved but rather smoothened-out in the $\Delta\Sigma$ curves.
The horizontal yellow dashed and purple dash-dotted lines in the top panels
represent the specific positions in the potential
as marked in Fig.~\ref{fig:Potential}, solid gray lines illustrate the
positions of the potential minima.
}
\label{fig:DynIrrevFixedDa}
\end{figure}
\begin{figure}
\centering
\includegraphics[scale=1]{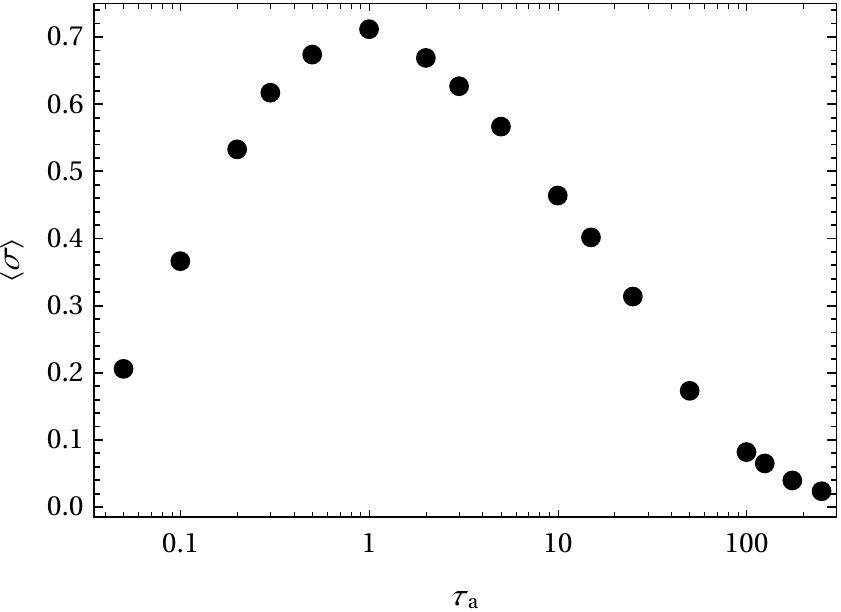}
\caption{Irreversibility production rate $\langle \sigma \rangle$
in the quartic double-well potential as a function of the
active correlation time $\ta$ for fixed active diffusion $\Da = 10$.
All other parameter values are as in Fig.~\ref{fig:DynIrrev}.}
\label{fig:IrrevRateFixedDa}
\end{figure}
We first consider decreasing $\ta$ for fixed $\Da = 10$.
Fig.~\ref{fig:DynIrrevFixedDa} shows trajectories for $\ta = 50$ (right),
$\ta = 1$ (middle) and $\ta=0.1$ (left) along with their irreversibility productions.
The production rate $\Delta\Sigma/\tau$ is larger for $\ta = 50$
than it is for $\ta = 250$ in Fig.~\ref{fig:DynIrrev}, and grows even further as $\ta = 1$.
At the same time the frequency of jumps increases as $\ta$ is lowered.
However, eventually the production rate decreases again ($\ta = 0.1$)
and is expected to reach zero in the white-noise limit $\ta \to 0$,
while the number of ``jumps'' continues to grow.
In this limit, diffusion is so strong that the potential barrier around
the origin affects the dynamics only marginally.
We remark that while $\Delta\Sigma$ is apparently negative in the passive white-noise limit,
the full entropy production is only obtained upon adding the system contributions
(the second term in \eref{eq:DeltaSigmaGen}) and must be zero in the steady state
since for passive Brownian motion the steady state corresponds to thermal equilibrium.
In our case, however, the total entropy production would assume a small positive
value from a short transient phase, because we are not starting from the
(joint) steady state and thus are slightly out of equilibrium in the beginning.

The dependence of the average irreversibility production rate
\begin{equation}
\label{eq:sigmaDef}
\langle\sigma\rangle = \lim_{\tau\to\infty} \langle\Delta\Sigma(\tau)\rangle/\tau
\end{equation}
on the active correlation time
is summarized in Fig.~\ref{fig:IrrevRateFixedDa}.
We can see the tendency towards vanishing irreversibility production rate
in the two limits $\ta \to \infty$ and $\ta \to 0$ as discussed above,
and a maximum of irreversibility production at around $\ta \approx 1$.
Surprisingly, the irreversibility production rate at the persistence time
$\ta = 250$, for which we can observe the distinct hysteresis-like behavior
and discern an ``arrow of time'' in the particle trajectories by naked eye
(cf.~Figs.~\ref{fig:DynIrrev} and \ref{fig:Propagator}),
is about a factor 20 smaller than the maximal rate at $\ta \approx 1$.
We might expect that this is due to a maximized rate of
``jumps'' between potential wells at $\ta \approx 1$, maximizing the number
of irreversible hysteresis-like cycles. Intuitively, a maximal ``jump'' rate
might occur when the persistence time $\ta$ of the active fluctuations is of the order of
the typical duration of a ``jump'' from one potential well to the other,
because then the next ``jump'' is initiated right after having finished the previous one.

\subsubsection{Decreasing active correlation time at constant active velocity.}
\begin{figure}
\centering
\includegraphics[scale=1.6]{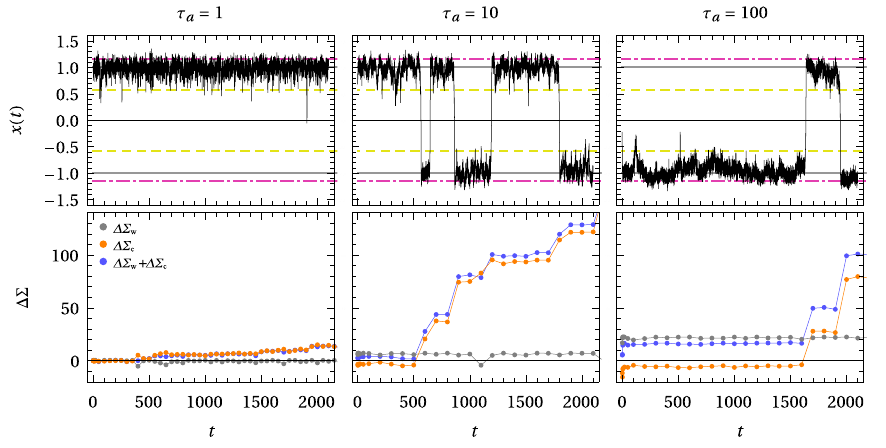}
\caption{Single trajectories (top) and corresponding irreversibility productions $\Delta\Sigma$
(bottom) for three systems
with different active correlation times
smaller than $\ta=250$ from Fig.~\ref{fig:DynIrrev}:
$\ta=1$ (left), $\ta=10$ (middle), and
$\ta = 100$ (right).
Here, the active velocity $v_{\mathrm{a}} = \sqrt{\Da/\ta} = 1/5$ is kept constant,
so that $\Da = 0.04$ (left), $\Da = 0.4$ (middle) and $\Da = 4.0$ (right), respectively.
All other parameter values are as in Fig.~\ref{fig:DynIrrev}.
In the lower panels, the dots mark the time points for which the values
of $\Delta\Sigma_{\mathrm{w}}$ (grey dots) and $\Delta\Sigma_{\mathrm{c}}$ (orange dots)
have been evaluated from the integrals in \eref{eq:DeltaSigmaw}
and \eref{eq:DeltaSigmac}; the connecting lines serve as a guide to the eye.
The time-resolution is considerably lower than for the trajectories in the upper panels.
The horizontal yellow dashed and purple dash-dotted lines in the top panels
represent the specific positions in the potential
as marked in Fig.~\ref{fig:Potential}, solid gray lines illustrate the
positions of the potential minima.
}
\label{fig:DynIrrevFixedva}
\end{figure}
\begin{figure}
\centering
\includegraphics[scale=1]{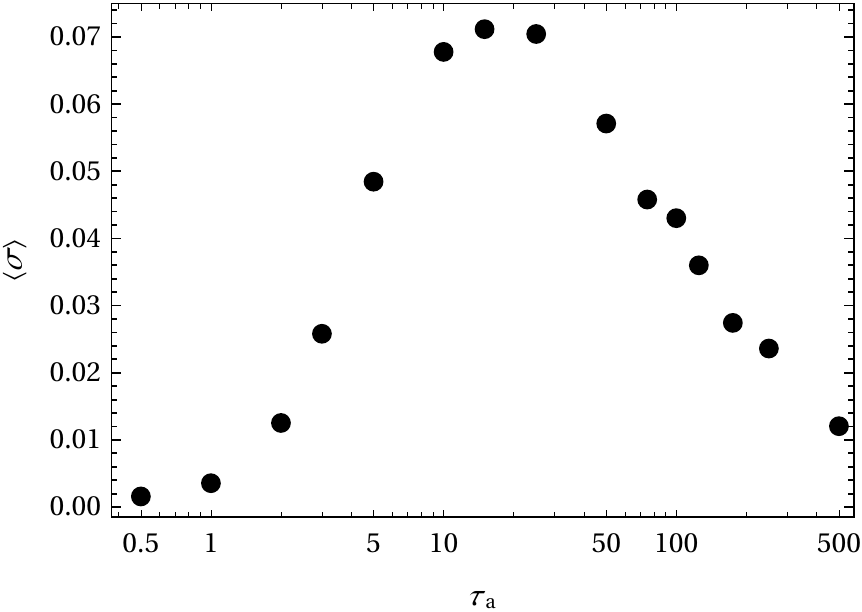}
\caption{Irreversibility production rate $\langle\sigma\rangle$
in the quartic double-well potential as a function of
the active correlation time $\ta$ for fixed active speed $v_{\mathrm{a}} = 1/5$.
All other parameter values are as in Fig.~\ref{fig:DynIrrev}.}
\label{fig:IrrevRateFixedva}
\end{figure}
The second option when taking the limit $\ta \to 0$ is to keep the self-propulsion
velocity $v_{\mathrm{a}} = \sqrt{\Da/\ta}$ of the particle fixed.
In the high-persistence case from Fig.~\ref{fig:DynIrrev} above,
its value is $v_{\mathrm a} = 1/5$.
Data for three correlation times smaller than $\ta=250$ are shown in Fig.~\ref{fig:DynIrrevFixedva}.
We observe that, compared to $\ta=250$ in Fig.~\ref{fig:DynIrrev},
the frequency of jumps first increases ($\ta=100$ and $\ta=10$) 
(note that the trajectories in Fig.~\ref{fig:DynIrrevFixedva} are about a factor 5
shorter than in Fig.~\ref{fig:DynIrrev}),
accompanied by an increase of irreversibility production.
Eventually, however, the jump frequency decreases again ($\ta=1$)
until jumps become very rare.
Summarizing, Fig.~\ref{fig:IrrevRateFixedva} shows the average irreversibility production
rate $\langle\sigma\rangle$ from \eref{eq:sigmaDef} as a function of $\ta$
(analogous to Fig.~\ref{fig:IrrevRateFixedDa}).
Again, we can see the irreversibility production rate approaching zero
in the limits $\ta \to \infty$ and $\ta \to 0$.
The maximal rate is reached at around $\ta \approx 10\ldots30$, and is roughly three times
as large as the rate at $\ta=250$, for which we have discussed the hysteresis-like
behavior of the particle trajectories and the associated irreversibility in
Sections~\ref{sec:firstNumerics} and \ref{sec:explaining} above.
Compared to the finding in Fig.~\ref{fig:IrrevRateFixedDa} for the
constant-$\Da$ scenario, the maximal irreversibility production rate
in Fig.~\ref{fig:IrrevRateFixedva} occurs
at about a $10$ times larger value for $\ta$. This observation is consistent with our intuitive
explanation of matching time-scales $\ta$ and typical ``jump times'',
because the latter is dominated by the active self-propulsion
velocity $v_{\mathrm{a}}$, which has the value $v_{\mathrm{a}}=\sqrt{10}$ at $\ta=1$ 
in Fig.~\ref{fig:IrrevRateFixedDa} and $v_{\mathrm{a}}=1/5$ in Fig.~\ref{fig:IrrevRateFixedva},
with roughly a factor 10 difference.

Finally, we would like to direct the reader's attention to the left panel of
Fig.~\ref{fig:DynIrrevFixedva}.
The particle is staying in one and the same potential well during the whole duration of the
trajectory, i.e.\ it is not jumping between potential wells, but $\Delta\Sigma$
increases nevertheless (though at a much slower rate than during the jumps).
It therefore seems that the steady state of an AOUP is irreversible even in a single-well
potential, provided the potential is anharmonic (cf.\ Section~\ref{sec:HP}).
In the following Section, we are going to briefly investigate
the irreversibility properties of an AOUP trapped in a quartic single-well potential.

\section{Quartic single-well potential}
\label{sec:QP}
\begin{figure}
\includegraphics[scale=0.85]{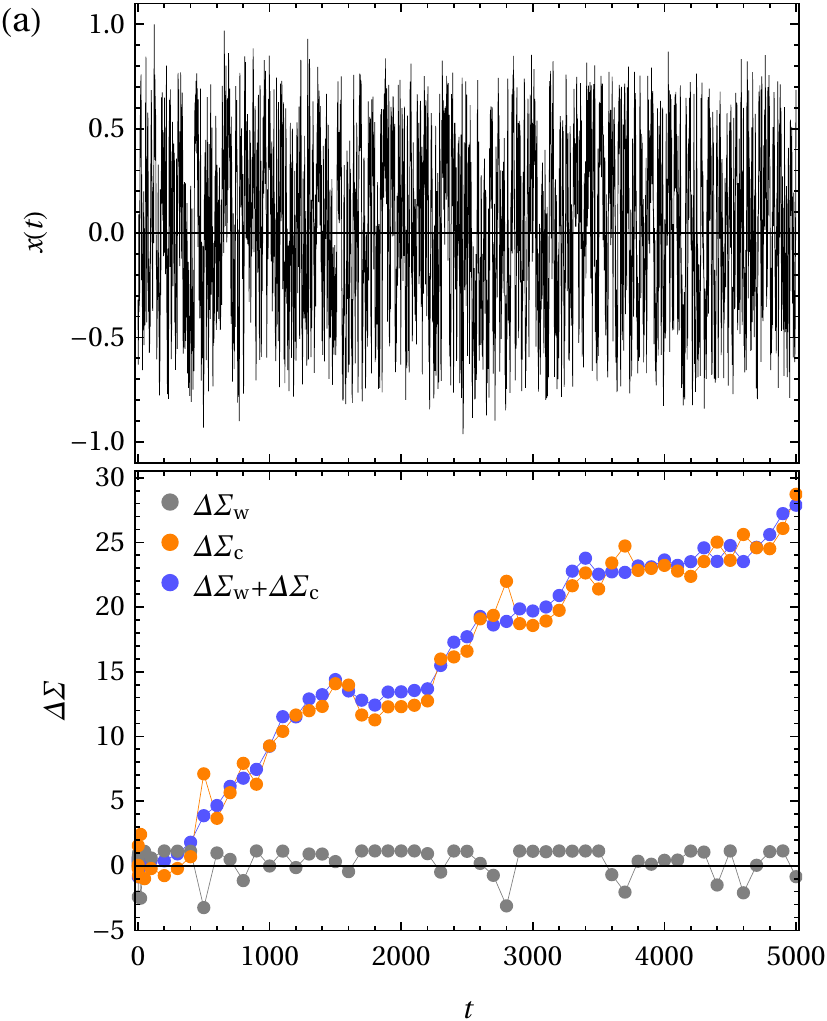}
\hspace*{0.5cm}
\includegraphics[scale=0.85]{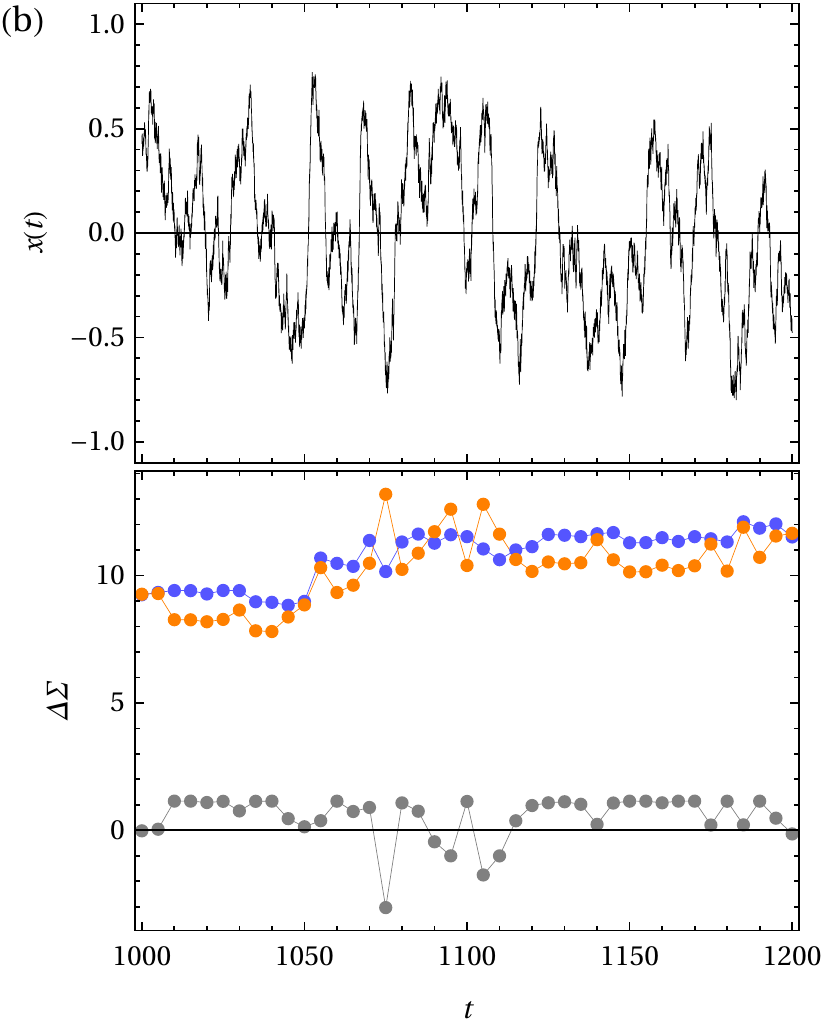}
\caption{(a) A single trajectory (top) and corresponding irreversibility productions $\Delta\Sigma$ (bottom)
in a quartic single-well potential ($k_4 = 1$, $k_2 = 0$) for the same dynamical parameters
as in Fig.~\ref{fig:DynIrrevFixedva} (left panels) 
(i.e.\  $\ta = 1$, $\Da = 0.04$, $D = 0.01$, $\gamma = 1$, $\kB=1$).
(b) Close-up of the same trajectory (top) and irreversibility productions (bottom) for
the time interval from about 1000 to 1200.
In the lower panels, the dots mark the time points for which the values of $\Delta\Sigma_{\mathrm{w}}$
and $\Delta\Sigma_{\mathrm{c}}$ have been evaluated from the integrals in \eref{eq:DeltaSigmaw}
and \eref{eq:DeltaSigmac}; the connecting lines serve as a guide to the eye.}
\label{fig:quartic1}
\end{figure}
\begin{figure}
\includegraphics[scale=0.85]{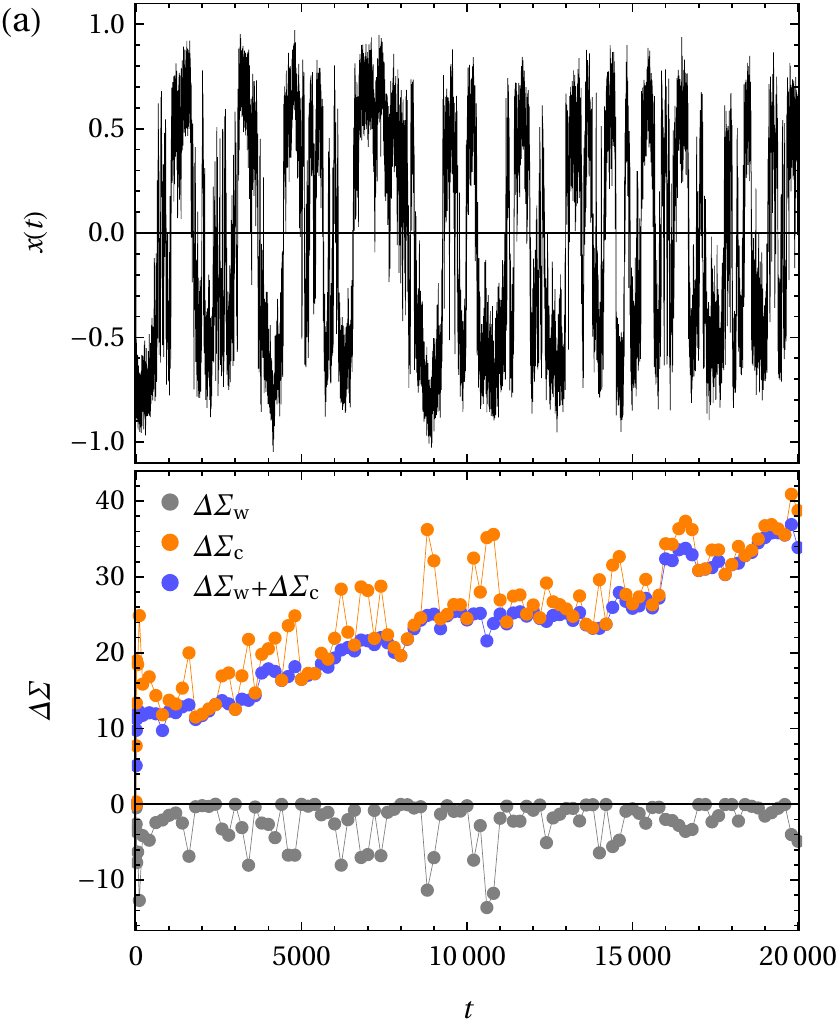}
\hspace*{0.5cm}
\includegraphics[scale=0.85]{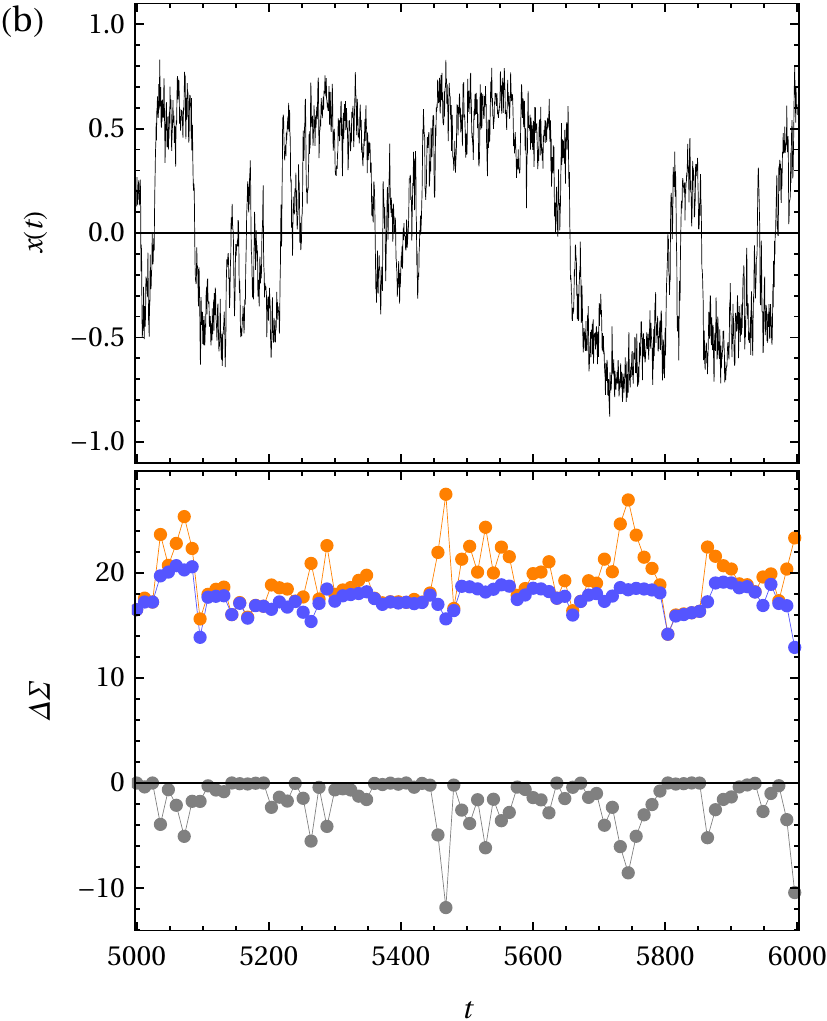}
\caption{(a) A single trajectory (top) and corresponding irreversibility productions $\Delta\Sigma$ (bottom)
in a quartic single-well potential ($k_4 = 1$, $k_2 = 0$) for the same dynamical parameters as
in Fig.~\ref{fig:DynIrrev} (i.e.\ $\ta = 250$, $\Da = 10$, $D = 0.01$, $\gamma = 1$, $\kB=1$).
(b) Close-up of the same trajectory (top) and irreversibility productions (bottom) for
the time interval from about 5000 to 6000.
In the lower panels, the dots mark the time points for which the values
of $\Delta\Sigma_{\mathrm{w}}$ (grey dots) and $\Delta\Sigma_{\mathrm{c}}$ (orange dots)
have been evaluated from the integrals in \eref{eq:DeltaSigmaw}
and \eref{eq:DeltaSigmac}; the connecting lines serve as a guide to the eye.}
\label{fig:quartic250}
\end{figure}
\begin{figure}
\includegraphics[scale=0.85]{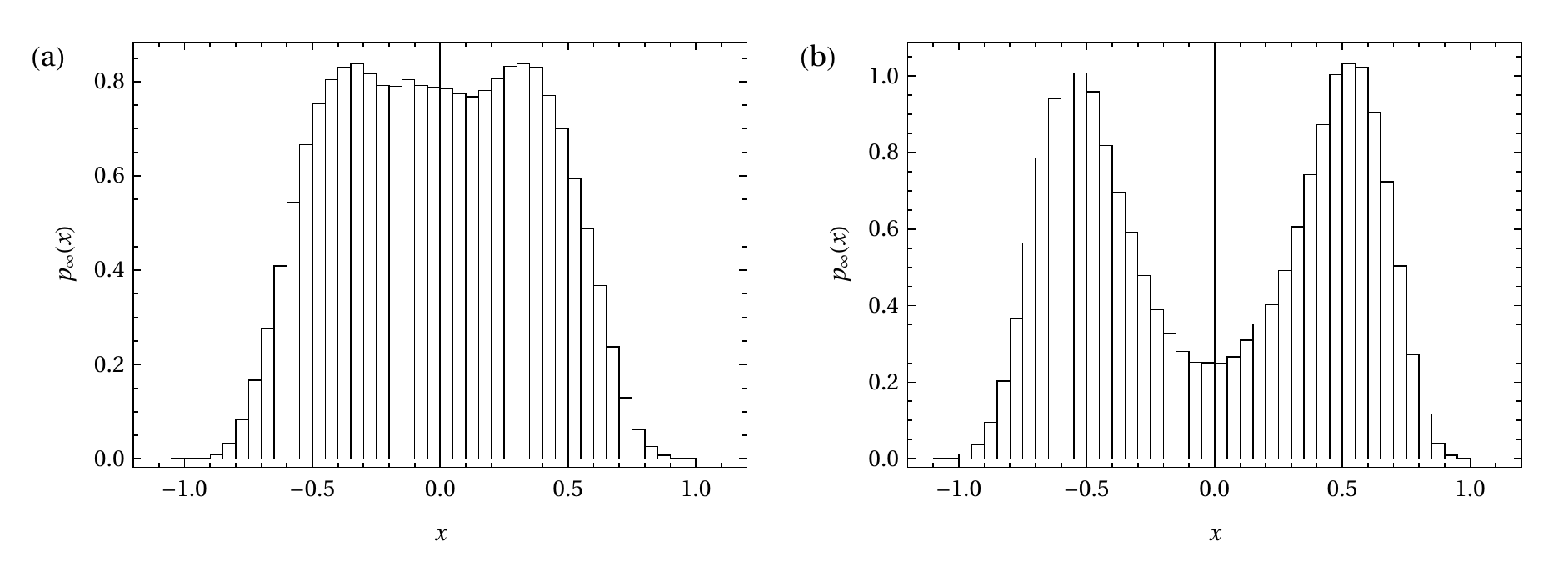}
\caption{Numerically computed steady-state distributions $p_\infty(x)$ of the AOUP trapped in a
quartic single-well potential ($k_4 = 1$, $k_2 = 0$ in \eref{eq:V}).
(a) Parameter values as in Fig.~\ref{fig:quartic1}
(i.e.\  $\ta = 1$, $\Da = 0.04$, $D = 0.01$, $\gamma = 1$, $\kB=1$).
(b) Parameter values as in Fig.~\ref{fig:quartic250}
(i.e.\ $\ta = 250$, $\Da = 10$, $D = 0.01$, $\gamma = 1$, $\kB=1$).
Deviations from perfect mirror symmetry are due to statistical fluctuations.}
\label{fig:steadyStateDistr}
\end{figure}
In this Section, we turn to the case of the AOUP \eref{eq:xLE}
moving in a quartic single-well potential with
$k_4>0$ and $k_2=0$ in \eref{eq:V}.
Figure \ref{fig:quartic1} shows a stationary-state trajectory (upper panels)
of the AOUP together with the
irreversibility production (lower panels)
for the same parameters of the active fluctuations as in the leftmost
panels of Fig.~\ref{fig:DynIrrevFixedva}.
Figure~\ref{fig:steadyStateDistr}(a) displays the corresponding steady-state distribution.
The effect of the active fluctuations is visible in the broadening of the distribution
(compared to the Boltzmann distribution it would assume without activity),
as the active fluctuations push the particle more towards the flanks of the potential. 
From Fig.~\ref{fig:quartic1}(a), we see
that the main contribution to irreversibility is
due to the colored-noise memory
kernel \eref{eq:DeltaSigmac} characteristic for the active fluctuations,
while the thermal white-noise component does not produce irreversibility.
Moreover, irreversibility seems to be produced  more or less continuously
without being connected to specific features of the trajectory
(like the ``jumps'' in the case of a double-well potential).
This observation is confirmed by
the closeup in Fig.~\ref{fig:quartic1}(b),
where it becomes apparent that
the transition-like movement of the particle from one
side of the potential well to the other
is not accompanied by a distinct production of irreversibility.
We can ``enhance'' these transitions by increasing the persistence time
of the active fluctuations, such that the active forcing $\eta(t)$ pushes the particle
to one side of the potential well for an extended period of time until $\eta(t)$
changes sign and drives the particle over to the other side.
We show a corresponding trajectory (upper panels) and the associated irreversibility
(lower panels) for $\ta=250$ in Fig.~\ref{fig:quartic250} (and with $\Da=10$,
i.e.\ the same active fluctuations for which we could observe the ``arrow of time''
in the double-well potential, see Fig.~\ref{fig:DynIrrev}).
The trajectory now shows ``jump-like'' transitions between the two edges of
the potential well at around $\pm 0.5$; see, in particular, the closeup in
Fig.~\ref{fig:quartic250}(b), and the associated steady-stated distribution in
Fig.~\ref{fig:steadyStateDistr}(b) with two peaks at the edges of the potential and
a depleted zone in the middle. However, there is
no clear indication that these ``jump-like'' transitions play a similar role
as a source of irreversibility like in the double-well potential.
In fact, the trajectory does not reveal any obvious features which would
tell us that it is considerably less likely to observe the same trajectory
in the stationary state but traced out backward in time.

From numerical simulations we find the maximal rate of irreversibility production
in the quartic single-well potential ($k_4=1$, $k_2=0$) when varying $\ta$
according to the constant $v_{\mathrm{a}}=\sqrt{\Da/\ta}$ scenario to be roughly 
$0.02$ at around $\ta = 5$.
Surprisingly, this maximal irreversibility production rate is only
about a factor 3 smaller than
the one in the double-well potential (see Fig.~\ref{fig:IrrevRateFixedva}),
even though in these two systems irreversibility is produced by
significantly different physical processes on distinct scales
(exploring the single quartic well vs.\ jumping between potential wells).
Despite these contrasts, the similar irreversibility productions in
the quartic single- and double-well potential imply that
the likelihood of observing time-reversed trajectories on average decreases at
comparable rates, and likewise the uncertainty when estimating the direction of the
arrow of time based on $\langle \sigma \rangle$ \cite{roldan2015decision}.

\section{Conclusions}
\label{sec:conclusions}
Active matter systems are inherently driven out of equilibrium as a result of their capability to
locally consume and convert energy (e.g., generating self-propulsion)
\cite{romanczuk2012active,elgeti2015physics,Bechinger:2016api,ramaswamy2017active,fodor2018statistical}.
Yet, the non-equilibrium nature of this active driving may not always be visible
or detectable in the emergent dynamical behavior of active particles,
in particular when the active system is being observed on the level of spatial
trajectories without resolving the microscopic processes generating the active
self-propulsion.
Active matter systems may therefore appear to bear certain equilibrium features 
despite their underlying non-equilibrium nature.
We here have assessed this potential resemblance by quantifying the irreversibility
of individual steady-state trajectories of active particles confined within a static
one-dimensional trapping potential. This setup excludes the occurrence of net (particle) currents
and of non-equilibrium stationary states sustained by external time-periodic driving forces.
Without activity, it corresponds to an equilibrium
situation with perfect path-wise reversibility.
To study the effect of activity on the (ir)reversibility properties of particle
trajectories we focused on the so-called active Ornstein-Uhlenbeck particle (AOUP)
as a ``minimal'', but popular and successful model for active Brownian motion
\cite{Fily:2012aps,Farage:2015eii, 
Maggi:2014gee,Argun:2016nbs,maggi2017memory, 
Fodor:2016hff,marconi2017heat,mandal2017entropy,Puglisi:2017crf,
koumakis2014directed,szamel2014self,szamel2015glassy,maggi2015multidimensional,Flenner:2016tng,paoluzzi2016critical,
Marconi:2016vdi,szamel2017evaluating,sandford2017pressure,wittmann2017effective,caprini2018linear,fodor2018statistical,bonilla2019active,
woillez2020nonlocal, 
berthier2013non,Marconi:2015tas,shankar2018hidden,dal2019linear,fodor2020dissipation}. 
In the AOUP model,
active self-propulsion is represented by a fluctuating colored-noise force
(Ornstein-Uhlenbeck process) in the equations of motion \cite{martin2020statistical},
see Section~\ref{sec:model}.

The most immediate
effect of activity is to create steady-state distributions within the trapping potential,
which are different from the equilibrium Boltzmann distribution
(see Fig.~\ref{fig:steadyStateDistr} for an example),
but which do not break any symmetry or carry any net currents.
Hence, intuition might have tempted us to expect the one-dimensional steady-state dynamics
to be path-wise reversible despite its active non-equilibrium character.
Our main results for the AOUP model are two-fold:

On the one hand, all steady-state trajectories are perfectly reversible in a harmonic potential,
exactly like in equilibrium. This result was proven analytically in Section \ref{sec:HP}.
It is valid for any trajectory of arbitrary duration and for any values of the system parameters.
In particular, it does not play any role whether the activity is weak or strong (controlled by $\Da$
in \eref{eq:xLE}) or whether the persistence time $\ta$ of the active driving is short or long.
Moreover, steady-state trajectories are reversible in a harmonic potential in arbitrary dimensions,
not just in a one-dimensional harmonic trap.
Our finding generalizes previous results regarding the reversibility of infinitely long trajectories of
a harmonically trapped AOUP \cite{dadhichi2018origins} to individual, finite-time trajectories and
corroborates earlier reports that an AOUP in a potential with vanishing third derivative
possesses equilibrium-like properties \cite{Fodor:2016hff,bonilla2019active,zamponi2005fluctuation,dal2019linear}.

On the other hand, AOUP trajectories are irreversible in a quartic potential, even in the steady state.
We demonstrated this by evaluating the irreversibility production
\eref{eq:DeltaSigmaGen} from numerically simulated trajectories
(which are much longer than transient relaxation processes).
In a quartic double-well potential (Section \ref{sec:DWP}) we identified
the jumps between the potential wells, driven by the active fluctuations, as the main source of irreversibility.
For large $\ta$, the irreversible nature of the individual particle trajectories
is even visible to the naked eye, such that we may quite easily distinguish
between trajectories being traced out forward in time versus backward in time.
In a quartic single-well potential, irreversibility does not seem to be connected
in such an obvious way to a distinct feature of the particle trajectories. Nevertheless, irreversibility
production can reach rates comparable to those in the double-well potential.
Moreover, in both the quartic single- and double-well potentials,
we find the irreversibility production to vanish, and thus the AOUP trajectories to
become reversible, for small and large $\ta$, in agreement with the fact that in these limits the
steady-state AOUP can be mapped to a passive Brownian particle at equilibrium
(see Appendix C in \cite{dabelow2019irreversibility} for an explicit computation
of the $\ta \to 0$ limit at constant $\Da$ for arbitrary potentials).

The associated decrease of the average irreversibility production rate $\langle\sigma\rangle$
with increasing $\ta$
is in contrast to the growing deviation of the steady-state distribution
from the equilibrium Boltzmann distribution
(see Fig.~\ref{fig:steadyStateDistr}).
This apparent contradiction (and others \cite{flenner2020active}) highlights that
the non-equilibrium characteristics of the steady state of active matter
are complex and subtle, in particular with respect to their analogy to
the equilibrium state.
Different hallmarks of equilibrium (Boltzmann distribution vs.\
path-wise reversibility vs.\ violations of Einstein relation etc.)
may capture different aspects of this analogy.
Concerning the log ratio of path probabilities $\Delta\Sigma$,
we would like to emphasize here that---by construction---it does not
measure entropy production in the thermal environment
and does not quantify departure from equilibrium in that sense.
In any model for active matter which simply represents activity by an ``effective active force''
(like the AOUP model) the entropy production and the corresponding departure from equilibrium
due to the microscopic dissipative processes generating the active self-propulsion drive
cannot be captured (see also the detailed discussion in \cite{dabelow2019irreversibility},
and \cite{pietzonka2017entropy} for an explicit model quantifying this microscopic entropy production).
Rather, $\Delta\Sigma$ assesses, in the very sense of its definition \eref{eq:DeltaSigmaDef},
how irreversible spatial AOUP trajectories are and how closely the AOUP dynamics
resembles the path-wise reversibility properties of equilibrium systems.

As we have seen in the present study, the AOUP dynamics can appear time reversible,
and thus equilibrium-like,
despite the non-equilibrium character of the active self-propulsion.
This result directly extends to other physical situations which are described by
our model \eref{eq:xLE}, \eref{eq:etaLE}. In particular, the equations of motion
\eref{eq:xLE}, \eref{eq:etaLE} are well
established to capture essential aspects of the dynamical behavior of
a passive tracer particle suspended in a ``bath'' of active swimmers
\cite{Maggi:2014gee,Argun:2016nbs,maggi2017memory,dabelow2019irreversibility}.
It would be interesting to connect our findings to the thermodynamic (-like)
properties of active matter \cite{takatori2016forces}, for instance,
by unravelling how the (ir)reversibility
of active particle trajectories is related to the active pressure
\cite{takatori2014swim,solon2015pressure} or to motility-induced phase
separation \cite{Cates:2015mip}.
Moreover, analogous studies should be carried out for other models of
active Brownian motion, like active Brownian particles or run-and-tumble particles
\cite{romanczuk2012active}.


\ack
LD acknowledges funding by the Deutsche Forschungsgemeinschaft (DFG)
within the Research Unit FOR 2692 under Grant No. 397303734
and by a Nordita Visiting PhD Student Fellowship.
SB wishes to thank Frank J\"ulicher, Charlie Duclut and Joris Paijmans for inspiring discussions.
RE acknowledges funding by the Swedish Research Council (Vetenskapsr{\aa}det)
under the Grants No.~2016-05412.
This work was supported by the Paderborn Center for Parallel Computing (PC$^2$)
within the Project ``hpc-prf-ubinmpt''.


\appendix
\section{Calculation of $\Gamma_{\mathrm{HP}}(t,t')$}
\label{app:GammaHP}
In eq.~\eref{eq:VHP} the explicit form of the ``memory kernel'' $\Gamma_{\mathrm{HP}}(t,t')$
in the path integral \eref{eq:PathWeightPositionHP}
is given for the case of an AOUP being trapped in a harmonic potential.
As mentioned in the main text, it turns up as the operator inverse
$\int_0^\tau \d t' \; V_{\mathrm{HP}}(t, t') \Gamma_{\mathrm{HP}}(t', t'') = \delta(t - t'')$
of the differential operator
\begin{eqnarray}
\label{eq:app:VHP}
\fl
V_{\mathrm{HP}}(t, t')
	= \delta(t - t') \left[
		- \ta^2 \partial_{t'}^2 + 1 + \frac{\Da}{D}
		+ \delta(t') \left( -\ta^2 \partial_{t'} - \ta + \bar{C}_{22} \right)
		+ \delta(\tau - t') \left(\ta^2 \partial_{t'} + \ta \right)
	\right]
\nonumber \\
\end{eqnarray}
when performing the Gaussian integral over all realizations $\Traj{\eta}$
of the active fluctuations.
Since the operator $V_{\mathrm{HP}}(t, t')$ is ``diagonal'' in its time arguments,
the integro-differential equation determining $\Gamma_{\mathrm{HP}}(t,t')$ simplifies
to the differential equation
\begin{eqnarray}
\label{eq:app:DGLGamma}
\fl
\left[
	- \ta^2 \partial_{t}^2 + 1 + \frac{\Da}{D}
	+ \delta(t) \left( -\ta^2 \partial_{t} - \ta + \bar{C}_{22} \right)
	+ \delta(\tau - t) \left(\ta^2 \partial_{t} + \ta \right)
\right] \Gamma_{\mathrm{HP}}(t, t') = \delta(t - t')
\, .
\nonumber \\
\end{eqnarray}
Similar equations have been solved
in the Appendices of \cite{dabelow2019irreversibility,dabelow2020},
investigating setups for AOUPs with initial conditions
different from the stationary state in a harmonic trap
considered here.
The mathematical procedure for finding the solution
of \eref{eq:app:DGLGamma} is essentially the same as in
\cite{dabelow2019irreversibility,dabelow2020}.
In fact, the differential equation (36) studied in the Appendix of \cite{dabelow2020}
is identical to our \eref{eq:app:DGLGamma} here when identifying
$1/\sigma^2$ with $\bar{C}_{22}$. We can therefore read off
$\Gamma_{\mathrm{HP}}(t,t')$ directly from the solution (17) in \cite{dabelow2020}
by setting $\sigma^2=1/\bar{C}_{22}$.

For the sake of completeness, we here briefly repeat the main ideas
and a few central steps of the calculation.
Exploiting that \eref{eq:app:DGLGamma} is a linear differential equation
we compose $\Gamma_{\mathrm{HP}}(t,t')$ from two parts,
$\Gamma_{\mathrm{HP}}(t,t') = G(t,t') + H(t,t')$.
The first part is the Green's function of the inhomogeneous equation
$[-\tau_a^2 \partial_t^2 + (1 + \Da/D) ] G(t, t') = \delta(t-t')$
with homogeneous boundary conditions $G(0, t') = G(\tau, t') = 0$,
the second part solves the homogeneous problem
$[-\tau_a^2 \partial_t^2 + (1 + \Da/D)] H(t,t') = 0$
such that the boundary terms are fixed as prescribed by \eref{eq:app:DGLGamma}.
We construct both parts, $G(t, t')$ and $H(t, t')$, from the general solution
\begin{equation}
\label{eq:app:Gamma}
\Gamma(t) = a^+ \e^{\lambda t} + a^- \e^{-\lambda t}
\, , \quad
\lambda = \frac{1}{\ta} \sqrt{1+\frac{\Da}{D}}
\,, \quad
a^\pm  = \mathrm{const}
\end{equation}
of the homogeneous ordinary differential equation
\begin{equation}
\left[ -\ta^2 \partial_{t}^2 + 1 + \frac{\Da}{D} \right]\Gamma(t) = 0
\end{equation}
associated with \eref{eq:app:DGLGamma}.

For $G(t,t')$, two such solutions, one for $0<t<t'$ and one for $t'<t<\tau$,
are matched at $t=t'$ such that the $\delta(t-t')$-inhomogeneity
appears when evaluating $[-\tau_a^2 \partial_t^2 + (1 + \Da/D) ] G(t, t')$.
The corresponding matching conditions at $t=t'$ and the homogeneous boundary
conditions $G(0, t') = G(\tau, t') = 0$ fix the two sets of parameters $a^\pm$
from the ansatz \eref{eq:app:Gamma} (one set for $0<t<t'$ and one for $t'<t<\tau$)
\cite{dabelow2019irreversibility}.
Via these matching conditions at $t=t'$, the time point $t'$ enters the solution
$G(t, t')$;
it is otherwise a fixed parameter in the differential equations
for $G(t,t')$ and $H(t,t')$, just like $D$, $\Da$, $\ta$ and $\bar{C}_{22}$.

For the function $H(t, t')$, we again make an ansatz of the form
\eref{eq:app:Gamma}.
We fix the coefficients $a^\pm$ by ensuring
that the full solution $\Gamma_{\mathrm{HP}}(t,t') = G(t,t')+H(t,t')$ with the
Green's function $G(t,t')$ already known from the previous step of the calculation
fulfills \eref{eq:app:DGLGamma}. Plugging $G(t,t')+H(t,t')$ into \eref{eq:app:DGLGamma},
and using $[-\tau_a^2 \partial_t^2 + (1 + \Da/D) ] G(t, t') = \delta(t-t')$
and $\left[ -\ta^2 \partial_{t}^2 + (1 + \Da/D) \right]H(t,t') = 0$,
we are left with the boundary contributions proportional to $\delta(t)$
and $\delta(\tau-t)$ on the left-hand side of \eref{eq:app:DGLGamma}
(expressed in terms of derivatives of $G(t,t')$,
the various system parameters, and the unknowns $a^\pm$)
and with zero on the right-hand side of \eref{eq:app:DGLGamma}.
Requiring that each of these boundary contributions vanishes,
we obtain $a^\pm$ for the function $H(t,t')$. 
Finally, the sought solution $\Gamma_{\mathrm{HP}}(t,t')$ is obtained according to
the superposition $\Gamma_{\mathrm{HP}}(t,t')=G(t,t')+H(t,t')$.

\section{Reversibility in the harmonic potential}
\label{app:DeltaSigmaHP=0}
The goal is to show that $\Delta\Sigma[\Traj{x}]$ as given in \eref{eq:DeltaSigmaHP} 
is identically zero.
We start by moving the time-derivatives within the integrals from $\dot{x}_t$
and $\dot{x}_{t'}$ to
$\Gamma_{\mathrm{HP}}(t,t')$ via partial integration and by sorting the resulting
expression for $\Delta\Sigma[\Traj{x}]$
into contributions containing genuine double time integrals, single time integrals
and pure boundary terms,
\begin{eqnarray}
\fl
\frac{\Delta\Sigma[\Traj{x}]}{\kB} =
\frac{1}{2} \int_0^\tau \d t \int_0^\tau \d t^\prime \, x_t K_2(t,t^\prime) x_{t^\prime}
+ \frac{1}{2} \int_0^\tau \d t \, x_t \left[ K_\tau(t) x_\tau -  K_0(t) x_0 \right]
+ \frac{1}{2} K \left( x_\tau^2 - x_0^2 \right)
\, ,
\nonumber \\
\label{eq:GammaHPsorted}
\end{eqnarray}
with
\begin{eqnarray}
K_2(t,t') & = &
\frac{\Da}{2D^2} 
\left[ 
  - \frac{4 k_2}{\gamma} \bar{\Gamma}_{\mathrm{HP}}^{(1,0)}(t, t^\prime)
  + \left(\frac{k_2}{\gamma}\right)^{\!\!2} \Delta\Gamma_{\mathrm{HP}}(t, t^\prime)
  + \Delta\Gamma_{\mathrm{HP}}^{(1,1)}(t, t^\prime)
\right]
\, ,
\nonumber \\
\label{eq:K2}
\\[2ex]
K_\tau(t) & = &
\frac{\Da}{2D^2} \left[
	\frac{4 k_2}{\gamma} \bar{\Gamma}_{\mathrm{HP}}(t,\tau)
	- 2 \Delta\Gamma_{\mathrm{HP}}^{(1,0)}(t,\tau)
\right]
\nonumber \\
&&\qquad \mbox{}
+ \frac{\sqrt{2\Da}}{D} \bar{C}_{12}\left[
	\frac{k_2}{\gamma}\Gamma_{\mathrm{HP}}(\tau-t, 0) - \Gamma_{\mathrm{HP}}^{(1,0)}(\tau-t, 0) 
\right]
\, ,
\label{eq:Ktau}
\\[2ex]
K_0(t) & = &
\frac{\Da}{2D^2} \left[
	\frac{4 k_2}{\gamma} \bar{\Gamma}_{\mathrm{HP}}(t,0)
	- 2 \Delta\Gamma_{\mathrm{HP}}^{(1,0)}(t,0)
\right]
\nonumber \\
&&\qquad \mbox{}
+ \frac{\sqrt{2\Da}}{D} \bar{C}_{12}\left[
	\frac{k_2}{\gamma}\Gamma_{\mathrm{HP}}(t, 0) - \Gamma_{\mathrm{HP}}^{(1,0)}(t, 0)
\right]
\, ,	
\label{eq:K0}
\\[2ex]
K & = &
\bar{C}_{11} - \left( \bar{C}_{12}^2 + \frac{\sqrt{2\Da}}{D}\bar{C}_{12}  \right) \, \Gamma_{\mathrm{HP}}(0,0)
- \frac{\Da}{2D^2} \Delta\Gamma_{\mathrm{HP}}(0,0)
- \frac{1}{D}\frac{k_2}{\gamma}
\, ,
\nonumber \\
\label{eq:K}
\end{eqnarray}
where we recall that
$\bar{\Gamma}_{\mathrm{HP}}
=\frac{1}{2}\left[ \Gamma_{\mathrm{HP}}(t, t^\prime)+\Gamma_{\mathrm{HP}}(\tau-t, \tau-t^\prime)\right]$
and
$\Delta\Gamma_{\mathrm{HP}}=\Gamma_{\mathrm{HP}}(t, t^\prime)-\Gamma_{\mathrm{HP}}(\tau-t, \tau-t^\prime)$,
as defined below eq.~\eref{eq:DeltaSigmaHP}.
To find these expressions we have exploited
$\Delta\Gamma_{\mathrm{HP}}(0,\tau)=\Delta\Gamma_{\mathrm{HP}}(\tau,0)=0$,
$\Delta\Gamma_{\mathrm{HP}}(\tau,\tau)=-\Delta\Gamma_{\mathrm{HP}}(0,0)$,
and we have used the symmetry
$\Gamma_{\mathrm{HP}}(t,t')=\Gamma_{\mathrm{HP}}(t',t)$, implying, e.g.,
$\Gamma_{\mathrm{HP}}(\tau,t')=\Gamma_{\mathrm{HP}}(t',\tau)$.
Moreover, we have introduced the notation
$\Gamma_{\mathrm{HP}}^{(i,j)}(t,t')$ to denote the $i$-th derivative
of $\Gamma_{\mathrm{HP}}(t,t')$ with respect to its first argument
and the $j$-th derivative with respect to its second argument,
an explicit example being
$\Gamma_{\mathrm{HP}}^{(1,0)}(\tau-t,0)
=\left. \frac{ \partial \Gamma_{\mathrm{HP}}(t, t') }{\partial t} \right|_{t = \tau - t, t' = 0}=\frac{\partial\Gamma_{\mathrm{HP}}(\tau-t,0)}{\partial(\tau-t)}
=-\frac{\partial\Gamma_{\mathrm{HP}}(\tau-t,0)}{\partial t}$.
Note that the contribution $K$ contains the $\delta(t-t')$-integral from
\eref{eq:DeltaSigmaHP} as a boundary term.
In the following we consider the four contributions
\eref{eq:K2}-\eref{eq:K} separately.

For convenience we briefly recall some central quantities.
We start with the abbreviations from \eref{eq:kappapmpm:def},
\begin{eqnarray}
\label{eq:kappapmpm}
\kappa_{\pm\pm} & = & 
\kappa_\pm \left( 1 - \kappa_\pm\ta/\bar{C}_{22} \right)
\nonumber \\
& = &
\kappa_\pm \left( 1 - \kappa_\pm\frac{D+\Da\rho^2}{2(D+\Da\rho)} \right)
\nonumber \\
& = &
\left( 1 \pm \lambda\ta \right)
	\left[ 1 - \left( 1 \pm \lambda\ta \right)\frac{D+\Da\rho^2}{2(D+\Da\rho)} \right]
\nonumber \\
& = & 
\left( 1 \pm \sqrt{1+\Da/D} \right)
	\left[ 1 - \left( 1 \pm \sqrt{1+\Da/D} \right)\frac{D+\Da\rho^2}{2(D+\Da\rho)} \right]
\, ,
\end{eqnarray}
where in the second line we have used the explicit form of $\bar{C}_{22}$
from \eref{eq:Cbar}, and in the third line the definition of
$\kappa_\pm = 1 \pm \lambda\ta = 1 \pm \sqrt{1+\Da/D}$
given in eq.~\eref{eq:kappapmpm:def}.
Some combinations of these constants, which we will need in the
following calculations, are
\begin{eqnarray}
\kappa_{++} + \kappa_{--} & = &
- \ta^2 \left[ \left(k_2/\gamma\right)^{\!2} + \lambda^2 \right] \frac{\Da \rho^2}{D+\Da\rho}
\label{eq:kpp+kmm}
\, ,
\\
\kappa_{++} - \kappa_{--} & = &
\ta^2 (2k_2\lambda/\gamma)\frac{\Da \rho^2}{D+\Da\rho}
\label{eq:kpp-kmm}
\, ,
\\
\kappa_{+-} - \kappa_{++} & = &
-\frac{1}{2} (\lambda\ta + 1) \left( \kappa_{++} - \kappa_{--} - 2\lambda\ta \right)
\nonumber \\
& = &
\lambda\ta \, (\lambda\ta + 1) \frac{D+\Da\rho^2}{D+\Da\rho}
\label{eq:kpm-kpp}
\, ,
\\
\kappa_{-+} - \kappa_{--} & = &
-\frac{1}{2} (\lambda\ta - 1) \left( \kappa_{++} - \kappa_{--} - 2\lambda\ta \right)
\nonumber \\
& = &
\lambda\ta \, (\lambda\ta - 1) \frac{D+\Da\rho^2}{D+\Da\rho}
\label{eq:kmp-kmm}
\, .
\end{eqnarray}
We also recall the explicit form of the integral kernel
$\Gamma_{\mathrm{HP}}(t,t')$ from \eref{eq:GammaHP},
\begin{eqnarray}
\label{eq:GammaHP:kappa}
\fl
\Gamma_{\mathrm{HP}}(t, t')
= \frac{\kappa_{+-} \e^{-\lambda |t - t'|} + \kappa_{-+} \e^{-\lambda(2\tau - |t - t'|)}
		- \kappa_{++} \e^{-\lambda(t+t')} - \kappa_{--} \e^{-\lambda(2\tau - t - t')}}
	   {2 \ta^2 \lambda \left( \kappa_{+-} - \kappa_{-+} \e^{-2 \lambda\tau} \right)}
\, ,
\end{eqnarray}
and the definitions
$\bar{\Gamma}_{\mathrm{HP}}=\frac{1}{2}\left[ \Gamma_{\mathrm{HP}}(t, t^\prime)+\Gamma_{\mathrm{HP}}(\tau-t, \tau-t^\prime)\right]$
and
$\Delta\Gamma_{\mathrm{HP}}=\Gamma_{\mathrm{HP}}(t, t^\prime)-\Gamma_{\mathrm{HP}}(\tau-t, \tau-t^\prime)$.

We can now compute the various combinations and derivatives of $\Gamma_{\mathrm{HP}}(t,t')$
appearing in \eref{eq:K2},
\begin{eqnarray}
\bar{\Gamma}_{\mathrm{HP}}^{(1,0)}(t, t^\prime)
& = &
\frac{\lambda\sign(t'-t) \left[ \kappa_{+-} \e^{-\lambda |t - t'|} - \kappa_{-+} \e^{-\lambda(2\tau - |t - t'|)} \right]}
	 {2 \ta^2 \lambda \left( \kappa_{+-} - \kappa_{-+} \e^{-2 \lambda\tau} \right)}
\nonumber \\
&& \qquad\quad \mbox{}
- \frac{\frac{\lambda}{2} \left( \kappa_{++} + \kappa_{--} \right) \left[ \e^{-\lambda(2\tau - t - t')} - \e^{-\lambda(t+t')} \right] }
       {2 \ta^2 \lambda \left( \kappa_{+-} - \kappa_{-+} \e^{-2 \lambda\tau} \right)}
\, ,
\label{eq:GammaHPbar(1,0)}
\\[2ex]
\Delta\Gamma_{\mathrm{HP}}(t, t^\prime)
& = &
\frac{\left( \kappa_{++} - \kappa_{--} \right) \left[ \e^{-\lambda(2\tau - t - t')} - \e^{-\lambda(t+t')} \right] }
       {2 \ta^2 \lambda \left( \kappa_{+-} - \kappa_{-+} \e^{-2 \lambda\tau} \right)}
\, ,
\\[2ex]
\Delta\Gamma_{\mathrm{HP}}^{(1,1)}(t, t^\prime)
& = &
\lambda^2 \, \Delta\Gamma_{\mathrm{HP}}(t, t^\prime)
\, .
\end{eqnarray}
The first line in \eref{eq:GammaHPbar(1,0)} is an odd function upon exchange of $t$ and $t'$ and
therefore does not contribute to the double integral in \eref{eq:GammaHPsorted}.
The remaining terms in $K_2(t,t')$ are proportional to
$\left[
(2k_2\lambda/\gamma) \left( \kappa_{++} + \kappa_{--} \right)
+ \left[ \left(k_2/\gamma\right)^{\!2} + \lambda^2 \right] \left( \kappa_{++} - \kappa_{--} \right)
\right] \left[ \e^{-\lambda(2\tau - t - t')} - \e^{-\lambda(t+t')} \right]$.
With \eref{eq:kpp+kmm} and \eref{eq:kpp-kmm} we see that
these terms are zero as well.

Next, we consider $K_\tau(t)$, see eq.~\eref{eq:Ktau}.
The expressions involving $\Gamma_{\mathrm{HP}}(t,t')$ read
\begin{eqnarray}
\fl
\bar{\Gamma}_{\mathrm{HP}}(t,\tau) =
\frac{\left[ \kappa_{+-} - \frac{1}{2}(\kappa_{++} + \kappa_{--}) \right] \e^{-\lambda(\tau-t)}
		+ \left[ \kappa_{-+} - \frac{1}{2}(\kappa_{++} + \kappa_{--}) \right] \e^{-\lambda(\tau+t)}}
	 {2 \ta^2 \lambda \left( \kappa_{+-} - \kappa_{-+} \e^{-2 \lambda\tau} \right)}
\, ,
\\[2ex]
\fl
\Delta\Gamma_{\mathrm{HP}}^{(1,0)}(t,\tau) =
\frac{\lambda(\kappa_{++} - \kappa_{--}) \left[ \e^{-\lambda(\tau-t)}+\e^{-\lambda(\tau+t)} \right]}
	 {2 \ta^2 \lambda \left( \kappa_{+-} - \kappa_{-+} \e^{-2 \lambda\tau} \right)}
\, ,
\\[2ex]
\fl
\Gamma_{\mathrm{HP}}(\tau-t, 0) =
\frac{\left( \kappa_{+-} - \kappa_{++} \right) \e^{-\lambda(\tau-t)}
		+ \left( \kappa_{-+} - \kappa_{--} \right) \e^{-\lambda(\tau+t)}}
	 {2 \ta^2 \lambda \left( \kappa_{+-} - \kappa_{-+} \e^{-2 \lambda\tau} \right)}
\, ,
\\[2ex]
\fl
\Gamma_{\mathrm{HP}}^{(1,0)}(\tau-t, 0) =
\frac{-\lambda \left( \kappa_{+-} - \kappa_{++} \right) \e^{-\lambda(\tau-t)}
		+ \lambda \left( \kappa_{-+} - \kappa_{--} \right) \e^{-\lambda(\tau+t)}}
	 {2 \ta^2 \lambda \left( \kappa_{+-} - \kappa_{-+} \e^{-2 \lambda\tau} \right)}
\, .
\end{eqnarray}
Plugging them into \eref{eq:Ktau},
using $\bar{C}_{12} = -\sqrt{2\Da}\frac{k_2\ta}{\gamma}\rho/(D+\Da\rho^2)$ (see \eref{eq:Cbar}),
and skipping common $t$-independent factors, we find
\begin{eqnarray}
K_\tau(t) & \propto &
\left\{
	\frac{2k_2\ta}{\gamma}\left[ \kappa_{+-} - \kappa_{++} + \frac{1}{2}(\kappa_{++} - \kappa_{--}) \right]
	- \lambda\ta(\kappa_{++} - \kappa_{--}) 
\right.
\nonumber \\
&& \qquad \mbox{}
\left.
	- \frac{2D\rho}{D+\Da\rho^2}\frac{k_2\ta}{\gamma}
		\left( \frac{k_2\ta}{\gamma}+\lambda\ta \right) \left( \kappa_{+-} - \kappa_{++} \right)
\right\}
\e^{-\lambda(\tau-t)}
\nonumber \\
& + & \left\{
	\frac{2k_2\ta}{\gamma}\left[ \kappa_{-+} - \kappa_{--} - \frac{1}{2}(\kappa_{++} - \kappa_{--}) \right]
	- \lambda\ta(\kappa_{++} - \kappa_{--})
\right.
\nonumber \\
&& \qquad \mbox{}
\left.
	- \frac{2D\rho}{D+\Da\rho^2}\frac{k_2\ta}{\gamma}
		\left( \frac{k_2\ta}{\gamma}-\lambda\ta \right) \left( \kappa_{-+} - \kappa_{--} \right)
\right\}
\e^{-\lambda(\tau+t)}
\nonumber \\
\end{eqnarray}
Using \eref{eq:kpp-kmm}, \eref{eq:kpm-kpp}, \eref{eq:kmp-kmm},
and keeping in mind that $1/\rho=1+k_2\ta/\gamma$ and $\Da/D=(\lambda\ta)^2-1$,
we recognize that the expressions in curly brackets in front of
$\e^{-\lambda(\tau-t)}$ and $\e^{-\lambda(\tau+t)}$
both vanish, such that we conclude $K_\tau(t)=0$.
By a completely analogous calculation we can show that $K_0(t)=0$ as well
(cf.~\eref{eq:K0}).

We finally turn to $K$, eq.~\eref{eq:K}.
With the expressions for $\bar{C}_{11}$ and $\bar{C}_{12}$ from \eref{eq:Cbar},
we can simplify
$\bar{C}_{11}-k_2/(D\gamma)=[k_2/(D\gamma)]\frac{-\Da\rho^2}{D+\Da\rho^2}$
and
$\bar{C}_{12}^2 + (\sqrt{2\Da}/D)\bar{C}_{12}=2\ta[k_2/(D\gamma)]\frac{-\Da\rho^2}{D+\Da\rho^2}\frac{D+\Da\rho}{D+\Da\rho^2}$.
Then, using
\begin{eqnarray}
\Gamma_{\mathrm{HP}}(0,0) & = &
\frac{(\kappa_{+-}-\kappa_{++}) + (\kappa_{-+}-\kappa_{--})\,\e^{-2\lambda\tau}}
	 {2 \ta^2 \lambda \left( \kappa_{+-} - \kappa_{-+} \e^{-2 \lambda\tau} \right)}
\, ,
\\
\Delta\Gamma_{\mathrm{HP}}(0,0) & = &
\frac{(\kappa_{++}-\kappa_{--}) \left( \e^{-2\lambda\tau}-1 \right)}
	 {2 \ta^2 \lambda \left( \kappa_{+-} - \kappa_{-+} \e^{-2 \lambda\tau} \right)}
\, ,
\end{eqnarray}
and eqs.~\eref{eq:kpp-kmm}, \eref{eq:kpm-kpp}, \eref{eq:kmp-kmm},
we rewrite $K$ as
\begin{eqnarray}
K & = &
\frac{k_2}{D\gamma}\frac{-\Da\rho^2}{D+\Da\rho^2}
\left[
1 - \frac{(\lambda\ta+1) + (\lambda\ta-1)\,\e^{-2\lambda\tau}}
	 	 {\kappa_{+-} - \kappa_{-+} \e^{-2 \lambda\tau}}
\right.
\nonumber \\
&&\qquad\qquad\qquad \mbox{}
\left.
+ \frac{\Da}{D}\frac{D+\Da\rho^2}{2(D+\Da\rho)}\frac{\e^{-2\lambda\tau}-1}
	 {\kappa_{+-} - \kappa_{-+} \e^{-2 \lambda\tau}}
\right]
\nonumber \\[2ex]
& = &
\frac{k_2}{D\gamma}\frac{-\Da\rho^2}{D+\Da\rho^2}
\left\{
1 -
\left[
	1 + \lambda\ta + \frac{\Da}{D}\frac{D+\Da\rho^2}{2(D+\Da\rho)}
\right]
\frac{1}{\kappa_{+-} - \kappa_{-+} \e^{-2 \lambda\tau}}
\right.
\nonumber \\
&&\qquad\qquad \mbox{}
\left.
+ \left[
	1 - \lambda\ta + \frac{\Da}{D}\frac{D+\Da\rho^2}{2(D+\Da\rho)}
\right]
\frac{\e^{-2 \lambda\tau}}{\kappa_{+-} - \kappa_{-+} \e^{-2 \lambda\tau}}
\right\}
\label{eq:Ksimplified}
\, .
\end{eqnarray}
Recalling that $\Da/D = (\lambda\ta)^2-1=-(1-\lambda\ta)(1+\lambda\ta)$
and comparing the two terms in the square brackets with \eref{eq:kappapmpm},
we can identify the square bracket in the first line
as $\kappa_{+-}$ and the square bracket in the second line as $\kappa_{-+}$.
We therefore conclude that $K=0$.

In summary, we hence find that $\Delta\Sigma[\Traj{x}]$ given
in \eref{eq:DeltaSigmaHP} vanishes identically, as claimed in the main text
(see Sec.~\ref{sec:HP}).

\bibliographystyle{iopart-num}
\bibliography{ms-arxiv-rev}

\end{document}